\documentclass[preprint,nofootinbib,floatfix,a4paper,prd,superscriptaddress]{revtex4-1}   	
\usepackage{geometry}                		
\usepackage[colorlinks,citecolor=blue]{hyperref}
\usepackage{amsmath}
\usepackage{mathrsfs}
\usepackage{bbm}
\usepackage{amsfonts}
\usepackage{amssymb}
\usepackage{latexsym}
\usepackage{graphicx}
\usepackage[english]{babel}
\usepackage{multirow}
\usepackage{float}
\usepackage{url}
\usepackage{slashed}
\usepackage{appendix}
\usepackage{orcidlink}

\newcommand{\be}{\begin{equation}}
\newcommand{\ee}{\end{equation}}
\newcommand{\ba}{\begin{array}}
\newcommand{\ea}{\end{array}}
\newcommand{\bea}{\begin{eqnarray}}
\newcommand{\eea}{\end{eqnarray}}

\usepackage{CormorantGaramond}

\def  \bcen   {\begin{center}}
\def  \ecen   {\end{center}}
\def  \beq    {\begin{equation}}
\def  \eeq    {\end{equation}}

\def\la   {\lambda}

\def\lee { \left( }
\def\rii { \right) }

\def\to {\rightarrow}

\newcommand{\TSUa}{\affiliation{\small Tsung-Dao Lee Institute $\&$ School of Physics and Astronomy, Shanghai Jiao Tong University, Shanghai 200240, China }}

\newcommand{\PU}{\affiliation{\small Phenikaa Institute for Advanced Study, Phenikaa University, Yen Nghia, Ha Dong, Hanoi 100000, Vietnam}}

\newcommand{\AS}{\affiliation{\small Institute of Physics, Academia Sinica, Nangang, Taipei 11529, Taiwan}}

\newcommand{\JN}
{\affiliation{\small 
Department of Physics, College of Physics $\&$ Optoelectronic Engineering,\\ 
Jinan University, Guangzhou 510632, P.R. China
}}

\begin{document}

\title{New Contributions to $b \to s \gamma $ in Minimal G2HDM}

\author{Che Hao Liu}
\email{chehao@gate.sinica.edu.tw}
\AS

\author{Van Que Tran \orcidlink{0000-0003-4643-4050}}
\email{vqtran@sjtu.edu.cn} \TSUa \PU

\author{Qiaoyi Wen
\orcidlink{0000-0001-5544-7123}}
\JN

\author{\\Fanrong Xu
\orcidlink{0000-0003-0986-8028}}
\email{fanrongxu@jnu.edu.cn}
\JN 

\author{Tzu-Chiang Yuan \orcidlink{0000-0001-8546-5031} \, }
\email{tcyuan@phys.sinica.edu.tw} \AS

\date{\today}							

\begin{abstract}

We study the flavor-changing bottom quark radiative decay $b \to s \gamma$ induced at one-loop level 
within the minimal gauged two-Higgs-doublet model (G2HDM). Among the three new contributions to this rare process  in G2HDM, we find that only the charged Higgs $\mathcal{H^\pm}$ contribution can be constrained by the current global fit data in $B$-physics. Other two contributions from the complex vectorial dark matter $\mathcal{W}$ and dark Higgs $\mathcal{D}$ are not sensitive to the current data.
Combining with theoretical constraints imposed on the scalar potential and electroweak precision data for the oblique parameters, we exclude mass regions $m_{\mathcal{H}^\pm} \lesssim 250$ GeV and $m_{\mathcal{D}} \lesssim 100$ GeV at the 95\% confidence level.

\end{abstract}

\maketitle


\section{Introduction\label{sec:Intro}}

The discovery of a Higgs boson near the vicinity of 125 GeV~\cite{ATLAS:2012yve,CMS:2012qbp} at the year 2012 completes the building blocks set up in the Standard Model (SM). Well-known unanswered questions in SM like the neutrino masses for neutrino oscillations, dark matter and dark energy problem for the cosmic energy reserve in the standard $\Lambda$CDM cosmology, gauge hierarchy problem concerning the stability of the electroweak scale under quantum fluctuations, {\it etc.} must be faced by new physics (NP) beyond the SM (BSM). With the direct search limits of new particles at the Large Hadron Collider (LHC) reach multi-TeV, many simple extensions of SM are either under severely constrained or completely ruled out. 
At this stage, it is upmost important to scrutinize a plethora of all available experimental data to explore where NP may still be hiding from us. Indirect probes of NP from loop-induced rare processes thus provide an unique opportunity in this endeavour. Rare $B$-meson decays can play a crucial role as both low-$p_T$ and high-$p_T$ searches at the LHCb and LHC respectively are accumulating more and more precise and complementary data for the indirect probes. 

In this paper, we focus on the one-loop process $b \to s \gamma$ in the minimal gauged two-Higgs-doublet model (G2HDM) advocated by some of us~\cite{Ramos:2021omo,Ramos:2021txu}.
The original model~\cite{Huang:2015wts} was motivated by gauging the popular inert two-Higgs-doublet model (I2HDM)~\cite{Barbieri:2006dq,LopezHonorez:2006gr,Arhrib:2013ela,Belyaev:2016lok,Tsai:2019eqi,Fan:2022dck} for scalar dark matter, augmented by an extended gauge-Higgs sector of $SU(2)_H \times U(1)_X$ with a hidden doublet and a hidden triplet. Thus the complete Higgs sector of the original model is quite rich but rather complicated to analyze. Nonetheless, various refinements~\cite{Arhrib:2018sbz,Huang:2019obt} and 
collider implications~\cite{Huang:2015rkj,Huang:2017bto,Chen:2018wjl,Chen:2019pnt,Dirgantara:2020lqy} were pursued with the same particle content as 
the original model. As demonstrated in~\cite{Ramos:2021omo,Ramos:2021txu,Tran:2022yrh}, one can drop the hidden triplet field of the extra $SU(2)_H$  without jeopardizing the symmetry breaking pattern and the mass spectra.
Furthermore omitting the hidden triplet vastly simplifies the scalar potential by getting rid of 6 parameters. We will refer this as minimal G2HDM, or simply G2HDM, in this work.

Since the two Higgs doublets $H_1$ and $H_2$ in I2HDM are lumped into an irreducible representation of the hidden $SU(2)_H$ in G2HDM, there are new Yukawa couplings between the SM and hidden heavy fermions with the inert Higgs doublet $H_2$. In fact, both the charged and neutral components of $H_2$ can couple one SM fermion in one generation and one hidden heavy fermion in another generation. The latter one gives rise to flavor changing neutral current (FCNC) Higgs interaction between one SM fermion and one hidden heavy fermion in different generations. 
Furthermore, unlike the extra gauge boson $W^{\prime \, \pm}$ in left-right symmetric models~\cite{Mohapatra:1979ia,Senjanovic:1975rk}, the $SU(2)_H$ gauge boson $\mathcal{W}^{(p,m)}$, one of the dark matter candidate in G2HDM, carries no electric charge and hence does not mix with the SM $W^\pm$.  $\mathcal{W}^{(p,m)}$ also give rise to FCNC gauge interaction via a right-handed current formed by one SM fermion and one hidden heavy fermion. 
All other neutral particles in G2HDM like the photon, $Z$, SM Higgs along with its hidden sibling as well as
the dark photon and dark $Z$ couple diagonally in flavors with the SM fermion pairs and heavy hidden fermion pairs.
Thus the naturalness of neutral current interactions proposed by Glashow and Weinberg~\cite{Glashow:1976nt} can be fulfilled in G2HDM as far as the SM sector is concerned.
Regarding this we note the following fine point: In~\cite{Glashow:1976nt}, a discrete $Z_2$ symmetry was 
imposed by hand in the scalar potential of the general 2HDM 
to forbid the unwanted FCNC Higgs interactions with SM fermions at tree level. In G2HDM, however, there is an accidental $h$-parity~\cite{Chen:2019pnt} in the model to guarantee the absence of SM particles couple to odd number of new particles with odd $h$-parity coming from the hidden sector.

All low energy FCNC processes must then be induced by quantum loops in G2HDM. 
This motivates our interests in rare $B$ meson decays, in particular $b \to s \gamma$ in this study.  We will focus on $b \to s \gamma$ in this work as a warm up and reserve the more complicated penguin process $b \to s l^+ l^-$ with $l = e$ or $\mu$ in our future effort.

The organization of this paper is as follows. In Section~\ref{sec:MG2HDM}, we give a succinct review of the minimal G2HDM. The relevant G2HDM interaction Lagrangians for the loop computations are given in Section~\ref{sec:G2HDMInteractions}, followed by a discussion of the Wilson coefficients that govern the amplitudes of $b \to s(\gamma,g)$ in Section~\ref{sec:WilsonCoeffs}. Relevant flavor phenomenology including renormalization group running effects is discussed in Section~\ref{sec:fpheno}.
In Section~\ref{sec:Numerics}, after a brief discussion of the scanning methodology we present our numerical results. We draw our conclusions in~\ref{sec:Conclusions}. Some analytical formulas are relegated to two appendices. Appendix~\ref{app:LoopAmps} gives the detailed expressions of the loop amplitudes entered in the Wilson coefficients, while Appendix~\ref{app:loopIntegrals} lists the Feynman parameterized loop integrals with all internal and external masses retained.


\section{Minimal G2HDM - A Succinct Review\label{sec:MG2HDM}}

In this Section, we will briefly review the minimal G2HDM. 
The quantum numbers of the matter particles in G2HDM under $SU(3)_C \times SU(2)_L \times SU(2)_H \times U(1)_Y \times U(1)_X$ 
are~\footnote{The last two entries in the tuples are the $Y$ hypercharge and $X$ charge of the two $U(1)$ factors. 
Note that the $Q_X$ charges of $\pm 1$  of the some fields in our earlier 
works~\cite{Huang:2015wts,Arhrib:2018sbz,Huang:2019obt,Chen:2018wjl,Huang:2017bto,Chen:2019pnt,Huang:2015rkj,Dirgantara:2020lqy} 
had been changed to $\pm 1/2$ here. This makes the interaction terms for the hidden $X$ gauge field look similar to those of the $B$ gauge field associated with the hypercharge. The anomaly cancellation 
remains intact with these changes.}

\noindent
\underline{Scalars}:
$$
H = \left( H_1 \;\; H_2 \right)^{\rm T}  \sim  \left( {\bf 1}, {\bf 2}, {\bf 2}, \frac{1}{2}, \frac{1}{2} \right) \; , $$
$$ \Phi_H  =\left( G^p_H \;\; \Phi^0_H \right)^{\rm T} \sim \left( {\bf 1}, {\bf 1}, {\bf 2}, 0, { \frac{1}{2} } \right) \; .
$$
We note that the two $SU(2)_L$ doublets $H_1$ and $H_2$ are grouped together as
$H=\left( H_1 \;\; H_2 \right)^{\rm T}$ to form a doublet of $SU(2)_H$ with $U(1)_X$ charge $+1/2$.

\noindent
\underline{Spin 1/2 Fermions}:

\underline{Quarks}
$$Q_L=\left( u_L \;\; d_L \right)^{\rm T} \sim \left(  {\bf 3}, {\bf 2}, {\bf 1}, \frac{1}{6}, 0 \right) \; , $$
$$U_R=\left( u_R \;\; u^H_R \right)^{\rm T} \sim  \left( {\bf 3}, {\bf 1}, {\bf 2}, \frac{2}{3}, {  \frac{1}{2} }  \right) \; , $$
$$D_R=\left( d^H_R \;\; d_R \right)^{\rm T} \sim \left( {\bf 3}, {\bf 1}, {\bf 2},  -\frac{1}{3}, {  - \frac{1}{2} }  \right) \; , $$
$$u_L^H \sim \left(  {\bf 3}, {\bf 1}, {\bf 1},  \frac{2}{3}, 0 \right) \; , \; d_L^H \sim \left(  {\bf 3}, {\bf 1}, {\bf 1}, -\frac{1}{3}, 0 \right) \; . $$

Even though the lepton sector is not relevant in this work, it is shown below for completeness. 

\underline{Leptons}
$$L_L=\left( \nu_L \;\; e_L \right)^{\rm T} \sim \left( {\bf 1}, {\bf 2}, {\bf 1},  -\frac{1}{2}, 0 \right) \; , $$ 
$$N_R=\left( \nu_R \;\; \nu^H_R \right)^{\rm T} \sim \left( {\bf 1}, {\bf 1}, {\bf 2},  0, {  \frac{1}{2} }  \right)  \; , $$
$$E_R=\left( e^H_R \;\; e_R \right)^{\rm T} \sim \left( {\bf 1}, {\bf 1}, {\bf 2},  -1, {  - \frac{1}{2} }  \right) \; , $$
$$\nu_L^H \sim \left( {\bf 1}, {\bf 1}, {\bf 1},  0, 0 \right) \; , \; e_L^H \sim \left( {\bf 1}, {\bf 1}, {\bf 1},  -1, 0 \right) \; .$$

The most general renormalizable Higgs potential which is invariant under both $SU(2)_L\times U(1)_Y$ and  $SU(2)_H \times  U(1)_X$  
can be written down as follows
\begin{align}\label{eq:V}
V = {}& - \mu^2_H   \left(H^{\alpha i}  H_{\alpha i} \right)
- \mu^2_{\Phi}   \Phi_H^\dag \Phi_H  
+  \lambda_H \left(H^{\alpha i}  H_{\alpha i} \right)^2   + \la_\Phi \lee \Phi_H^\dag \Phi_H  \rii^2  \nonumber \\
{}& + \frac{1}{2} \lambda'_H \epsilon_{\alpha \beta} \epsilon^{\gamma \delta}
\left(H^{ \alpha i}  H_{\gamma  i} \right)  \left(H^{ \beta j}  H_{\delta j} \right)  \\
{}&
+\lambda_{H\Phi} \lee H^\dag H  \rii  \lee \Phi_H^\dag \Phi_H \rii  
 + \lambda^\prime_{H\Phi} \lee H^\dag \Phi_H  \rii  \lee \Phi_H^\dag H \rii,  \nonumber
\end{align}
where  ($i$, $j$)  and ($\alpha$, $\beta$, $\gamma$, $\delta$) refer to the $SU(2)_L$ and $SU(2)_H$ indices respectively, 
all of which run from one to two. We denote $H^{\alpha i} = H^*_{\alpha i}$, so $H^\dagger H = H^{\alpha i}  H_{\alpha i}$ and 
$\left( H^\dag \Phi_H \right) \left( \Phi_H^\dagger H \right) = \left( H^{\alpha i} \Phi_{H \alpha}\right) \left( \Phi^*_{H \beta} H_{\beta i} \right)$. 

To study spontaneous symmetry breaking (SSB) in the model, we parameterize the Higgs fields linearly according to standard lore
\begin{eqnarray}
\label{eq:scalarfields}
H_1 = 
\begin{pmatrix}
G^+ \\ \frac{ v + h_{\rm SM}}{\sqrt 2} + i \frac{G^0}{\sqrt 2}
\end{pmatrix}
\; , \;\;\;
H_2 = 
\begin{pmatrix}
 \mathcal H^+  \\  \mathcal H_2^0 
\end{pmatrix}
\; , 
\end{eqnarray}
\begin{eqnarray}
\Phi_H = 
\begin{pmatrix}
G_H^p  \\ \frac{ v_\Phi + \phi_H}{\sqrt 2} + i \frac{G_H^0}{\sqrt 2}
\end{pmatrix}
\;  ,
\end{eqnarray}
where $v$ and $v_\Phi$ are the only non-vanishing vacuum expectation values (VEVs)
in the SM doublet $H_1$ and the hidden doublet $\Phi_{H}$ fields respectively, with  $v = 246$ GeV is the SM VEV and
$v_\Phi$ a hidden VEV at the TeV scale. $H_2$ is the inert doublet with $\langle H_2 \rangle = 0$. 
In essence, the scalar sector of minimal G2HDM is a special tailored 3HDM.

\section{G2HDM Interactions\label{sec:G2HDMInteractions}}

In this Section, we provide the relevant interaction Lagrangians and other information for the computation of $b \to s (\gamma, g)$ at one-loop in minimal G2HDM. We will mainly follow the convention in Peskin and Schroeder~\footnote{M.~E.~Peskin and D.~V.~Schroeder,
``{\it An Introduction to quantum field theory},"
Addison-Wesley, 1995,
ISBN 978-0-201-50397-5}.

Besides introducing the CKM unitary mixing matrix 
\begin{equation}
\label{VCKM}
V_{\rm CKM} \equiv \left( U_u^L \right)^\dagger U^L_d  = 
\begin{pmatrix}
V_{ud} & V_{us} &V_{ub} \\
V_{cd} & V_{cs} &V_{cb} \\
V_{td} & V_{ts} &V_{tb}
    \end{pmatrix} \; ,
\end{equation}
while one diagonalizes the mass matrices of SM quarks, we also need to introduce the following two unitary mixing matrices while one diagonalizes the mass matrices of heavy new quarks in G2HDM,
\begin{align}
\label{VHu}
V^H_u & \equiv \left( U^R_u \right)^\dagger U^R_{u^H} \, , \\
\label{VHd}
V_d^{H} & \equiv \left( U^R_d \right)^\dagger U^R_{d^H} \; .
\end{align}

\subsection{Photon, Gluon and $W^\pm$ Interactions\label{PhotonandGluonInt}}

For the photon, the relevant interaction Lagrangian is
\bea
\label{Lgamma}
\mathcal L^\gamma  &  \supset  & - i e \left(  \mathcal H^+ \stackrel{\leftrightarrow}{\partial}_\mu \mathcal H^- \right) A^\mu  \nonumber \\
&& + \, e \biggl[ Q_u \sum_{q=u,c,t} \left( \bar q \gamma_\mu q + \overline{q^H} \gamma_\mu q^H \right)  + Q_d \sum_{q=d,s,b} \left( \bar q \gamma_\mu q + \overline{q^H} \gamma_\mu q^H \right) \biggr] A^\mu \nonumber \\
&& + \, i e \biggl[ \left( \partial_\mu W^+_\nu - \partial_\nu W^+_\mu \right) W^{\mu -} A^\nu - \left( \partial_\mu W^-_\nu - \partial_\nu W^-_\mu \right) W^{\mu +} A^\nu \biggr. \nonumber \\
&& \;\; \;\; \biggl. + \frac{1}{2} \left( \partial_\mu A_\nu - \partial_\nu A_\mu \right) \left( W^{\mu +}W^{-\nu} - W^{-\mu} W^{+\nu} \right) \biggr] \; ,
\eea
where $( a \stackrel{\leftrightarrow}{\partial}_\mu b ) \equiv a \partial_\mu b - b \partial_\mu a$, $Q_u=2/3$ and $Q_d=-1/3$.

For the gluon, we have
\bea
\label{Lgluon}
\mathcal L^g  &  \supset  & g_s \sum_{q=u,d,s,c,b,t} \left(  \bar q T^a \gamma_\mu   q + \overline{q^H} T^a \gamma_\mu   q^H \right) G^\mu_a \; ,
\eea
where $T^a$ are the generators of the color group $SU(3)_C$ associated with the gluon fields $G^\mu_a$ for $a=1,\cdots,8$.

\begin{figure}[tb]
         \centering
	\includegraphics[width=1.0\textwidth]{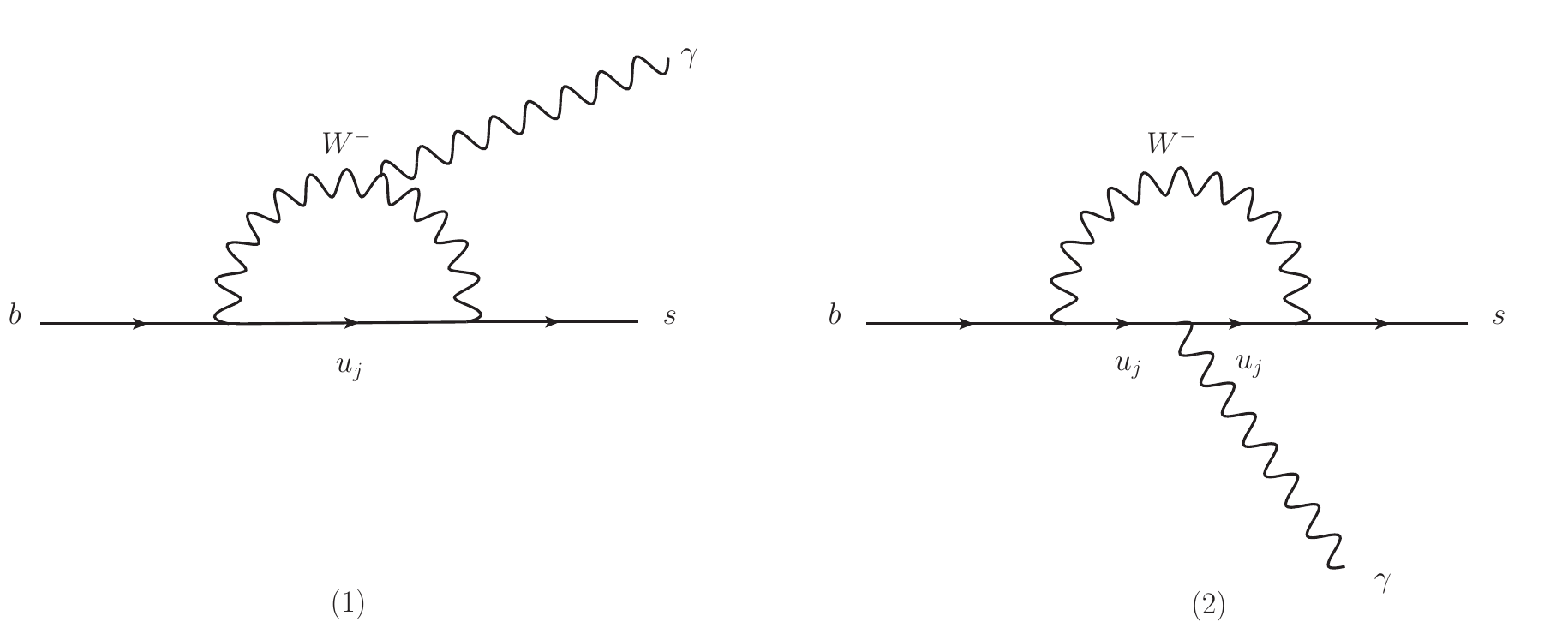}
	\caption{ \label{fig:bsgammaW} Contributions to $b \to s \gamma$ from the SM $W^\pm$ loop in the unitary gauge.}
\end{figure}

The SM charged current interaction for the quarks is
\beq
\label{LW}
\mathcal L^W \supset \frac{g}{2 \sqrt 2} \sum_{i,j} \bar u_j \left( V_{\rm CKM} \right)_{ji} \gamma^\mu \left( 1 - \gamma_5 \right) d_i W^{+}_\mu + {\rm h.c.} 
\eeq
where $V_{\rm CKM}$ is defined in (\ref{VCKM}) with $i,j$ being the generation indices. Since the effective Lagrangian describing the rare FCNC decays $b \to s(\gamma,g)$ is given by the chirality flipped transition dipole operators, the chiral $V-A$ structure of SM interaction (\ref{LW}) implies the loop amplitudes can enjoy the enhancement by two internal top quark mass insertions, besides the mass insertion from either side of the external lines due to equation of motion. This is to be compared with the similar processes $t \to c (\gamma, g)$ in which SM contribution arises from the hermitian conjugate of (\ref{LW}), and hence involves two bottom quark mass insertions instead. This distinctive feature is reflected in the SM branching ratios of $b \to s \gamma$ and $t \to c \gamma$ which are about $\sim 3\times 10^{-4}$ \cite{Misiak:2006zs, Belle:2014nmp} and $\sim 10^{-14}$ \cite{Aguilar-Saavedra:2002lwv} respectively. 

\subsection{G2HDM Interactions\label{CCInts}}

There are three new charged (electric charge or dark charge) current interactions in G2HDM 
mediated by the dark Higgs $\mathcal D$, charged Higgs $\mathcal H^\pm$, and $\mathcal W^{(p,m)}$ that can give rise to $b \to s (\gamma, g)$ at one-loop.
The first contribution is from the dark Higgs $\mathcal D$ which is a linear combination of two odd 
$h$-parity components $\mathcal H_2^{0}$ and $G_H^m$~\footnote{The other orthogonal combination is $\tilde G = -\sin \theta_2 \mathcal H_2^{0} + \cos \theta_2 G_H^m$, which together with its complex conjugate, are the Goldstone bosons absorbed by the longitudinal components of $\mathcal W^{(m,p)}$.}
\beq
\mathcal D = \cos\theta_2 \mathcal H_2^{0} + \sin\theta_2 G_H^m \; ,
\eeq
where $\theta_2$ is a mixing angle giving by
\beq
\tan 2 \theta_2 = \frac{2 v v_\Phi}{v_\Phi^2 - v^2} \; .
\eeq
The mass of $\mathcal D$ is
\beq
m_{\mathcal D}^2 = \frac{1}{2} \lambda^\prime_{H \Phi} \left( v^2 + v_\Phi^2 \right) \; . 
\eeq

\begin{figure}[tb]
         \centering
	\includegraphics[width=0.6\textwidth]{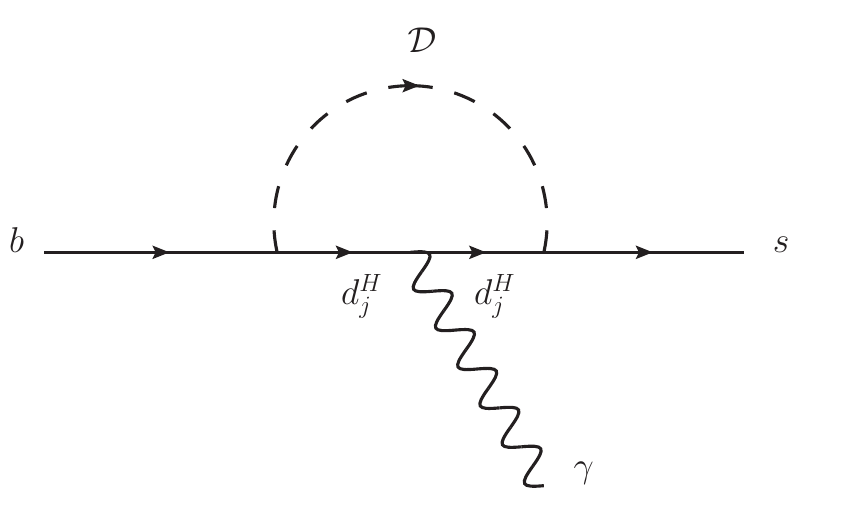}
	\caption{ \label{fig:bsgammaD} Contribution to $b \to s \gamma$ from the $\mathcal{D}$ loop.}
\end{figure}

The relevant interaction Lagrangian for the dark boson $\mathcal D$ interacts with the SM down-type quarks $d_i$ and new heavy down-type quarks $d^H_j$ 
in G2HDM is given by
\bea
\label{LD}
{\mathcal L}^{\mathcal D} & \supset & 
\sum_{i, j}  \overline{ d^H_j } \left[ \left( S^{\mathcal D}_{d} \right)_{ji} + \left( P^{\mathcal D}_{d} \right)_{ji} \gamma_5 \right] d_i \, \mathcal D^* + \; {\rm h.c.} 
\eea
where the Yukawa couplings matrices $S^{\mathcal D}_{d}$ and $P^{\mathcal D}_{d}$ are given by
\bea
\label{YukDS}
( S^{\mathcal D}_{d} )_{ji} & = &  \frac{\sqrt 2}{2 v}  \cos \theta_2 \left( V_d^{H \, \dagger} M_{d}  \right)_{ji} + 
\frac{\sqrt 2}{2 v_\Phi}   \sin \theta_2  \left( M_{d^H} V_d^{H \, \dagger} \right)_{ji} \; , \\
\label{YukDP}
( P^{\mathcal D}_{d} )_{ji} & = &  - \frac{\sqrt 2}{2 v}  \cos \theta_2 \left( V_d^{H \, \dagger} M_{d}  \right)_{ji} + 
\frac{\sqrt 2}{2 v_\Phi}   \sin \theta_2  \left( M_{d^H} V_d^{H \, \dagger} \right)_{ji} \; , 
\eea
with $V^H_{d}$ defined in (\ref{VHd}) and
\bea
M_d & = & {\rm diag} \left( m_d, m_s, m_b \right) \; , \\
M_{d^H} & = & {\rm diag} \left( m_{d^H}, m_{s^H}, m_{b^H} \right) \; .
\eea
Note that the ordering of the mass matrices and the mixing matrices are important in the Yukawa couplings (\ref{YukDS}) and (\ref{YukDP}). Also, 
fixing $v$, $v_\Phi$ and $V_d^H$,  for small (large) mixing angle $\theta_2$, these Yukawa couplings are suppressed (enhanced) by the down-type quark (heavy quark) mass $M_d$ ($M_{d^H}$). For $v_\Phi \gg v$, the contributions from $\mathcal D$ 
are expected to be minuscule. Similar small effects from $\mathcal D$ was found in 
$t\to c(\gamma,g)$ as well~\cite{tcgamma}.

\begin{figure}[tb]
         \centering
	\includegraphics[width=1.0\textwidth]{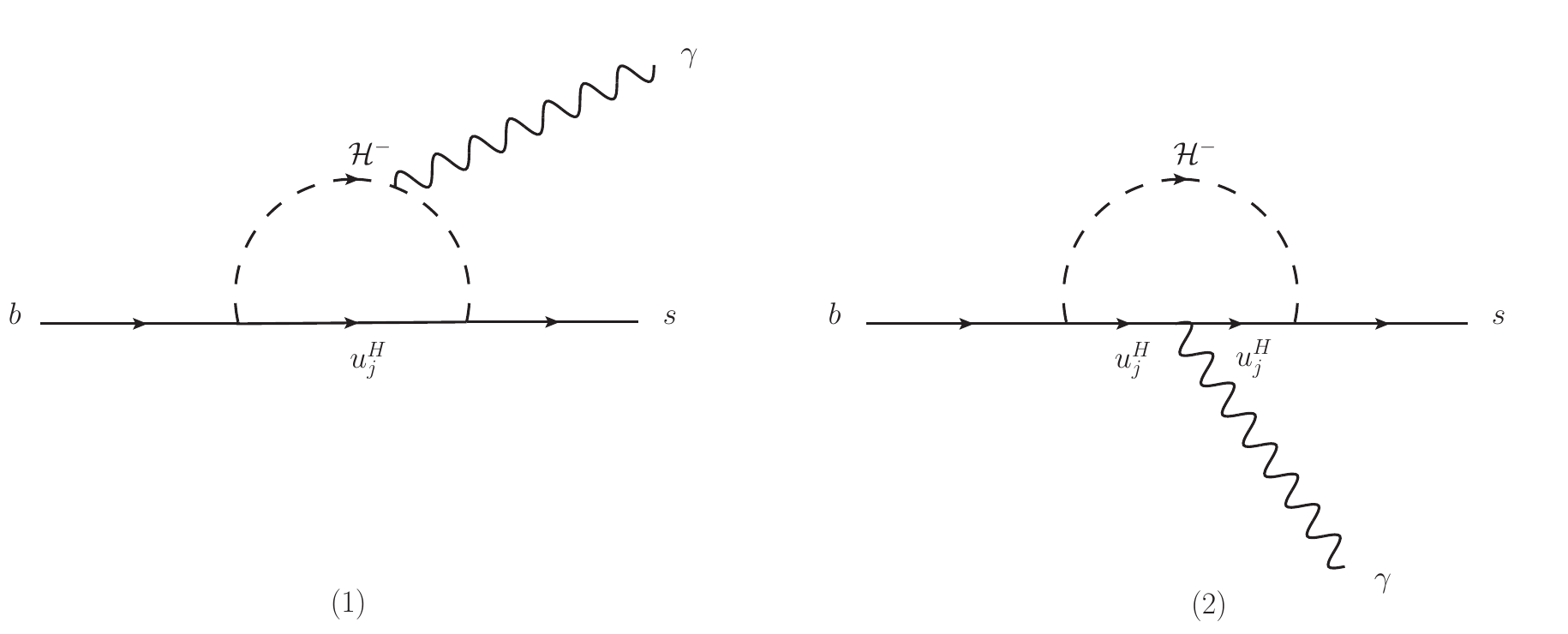}
	\caption{ \label{fig:bsgammaH} Contributions to $b \to s \gamma$ from the $\mathcal{H^\pm}$ loop.}
\end{figure}
The second contribution to $b \to s (\gamma, g)$ is from the dark charged Higgs $\mathcal H^\pm$ which is quite peculiar in G2HDM as 
compared with other multi-Higgs doublet model since it has odd $h$-parity. Thus the following vertices 
$W^\pm \mathcal H^\mp \gamma$, $W^\pm \mathcal H^\mp Z$ and $W^\pm \mathcal H^\mp h$ 
are all nil in the model. The mass of the charged Higgs is given by 
\beq
m_{\mathcal{H}^\pm}^2 = \frac{1}{2} \left( \lambda^\prime_{H\Phi} v_\Phi^2  - \lambda^\prime_H v^2 \right) \; .
\eeq
The relevant interaction Lagrangian for the charged Higgs exchange is 
\beq
\label{LChHiggs}
{\mathcal L}^{\mathcal H} \supset 
 \sum_{i, j}  \overline{ u^H_j } \left[ \left( y^{\mathcal H}_{u} \right)_{ji} \left( 1 - \gamma_5 \right) \right] d_i \, \mathcal H^+ + \; {\rm h.c.} 
\eeq
where the  Yukawa coupling matrix $y^{\mathcal H}_{u}$ is given by 
\bea
\label{YukChHd}
( y^{\mathcal H}_{u})_{ji}   =  \frac{\sqrt 2}{2 v} \left(  V_u^{H \, \dagger}  M_u  V_{\rm CKM} \right)_{ji} \; ,
\eea
with $V_{\rm CKM}$ and $V^H_{u}$ defined in (\ref{VCKM}) and (\ref{VHu}) respectively, and
\beq
M_u  =  {\rm diag} \left( m_u, m_c, m_t \right) \; .
\eeq
Since the Yukawa coupling $y^{\mathcal H}_u$ is proportional to the up-type quark mass matrix $M_u$, we expect charged Higgs contribution to $b \to s (\gamma,g)$ from the third generation heavy fermions is more relevant than the $\mathcal D$ contribution. This is to be compared with the charged Higgs contribution to $t \to c (\gamma, g)$ where the 
corresponding Yukawa coupling $y^\mathcal{H}_d = \sqrt 2 (V_{\rm CKM} M_d V^{H}_d) / 2 v$ is proportional to the down-type quark mass matrix $M_d$ and therefore has smaller impact~\cite{tcgamma}.

The third contribution to $b \to s (\gamma, g)$ is from the vector dark matter $\mathcal W^{(p,m)}$
assumed to be the lightest $h$-parity odd particle in minimal G2HDM with mass given by 
\beq
m_{\mathcal{W}} = \frac{1}{2} g_H \sqrt{ v^2 + v_\Phi^2 } \; .
\eeq
The relevant interaction Lagrangian for $\mathcal{W}$ is given by
\beq
\label{LWp}
{\mathcal L}^{\mathcal W} \supset \frac{g_H}{2 \sqrt 2} \sum_{i,j}  \overline{ d^H_j }  \left[    \left( V^{H \, \dagger}_d \right)_{ji} \gamma^\mu ( 1 + \gamma_5 ) \right]  d_i \mathcal  W^p_\mu  + {\rm h.c.} 
\eeq
It is interesting to note that the dark matter gauge boson $\mathcal{W}$ couples to a right-handed current formed by one SM fermion and one hidden heavy fermion. However from our previous works, we know the hidden gauge coupling $g_H$ is constrained to be small, of order 
one percent or less, we expect the contribution to the processes $b \to s (\gamma, g)$ from 
the dark matter $\mathcal W$ is not significant too. 
Similar situation is found in $t \to c (\gamma,g)$~\cite{tcgamma}.

\begin{figure}[tb]
         \centering
	\includegraphics[width=0.6\textwidth]{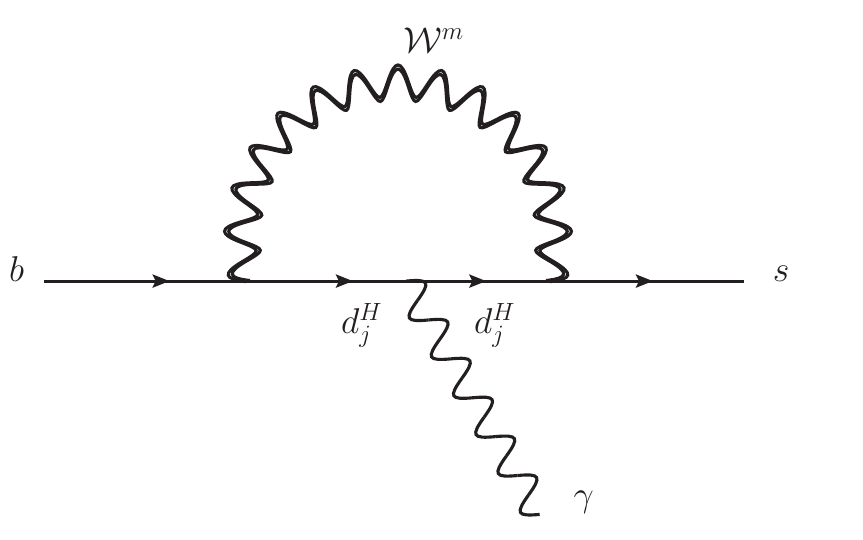}
	\caption{ \label{fig:bsgammaWprime} Contribution to $b \to s \gamma$ from the $\mathcal{W}^{(p,m)}$ loop in the unitary gauge.}
\end{figure}

The interaction Lagrangians $\mathcal L^{\mathcal D}$, $\mathcal L^{\mathcal H}$ and $\mathcal L^{\mathcal W}$ given by (\ref{LD}), (\ref{LChHiggs}) and (\ref{LWp}) respectively,
are the three new contributions from minimal G2HDM that can induce one-loop FCNC  
$b \to s (\gamma , g)$ decays competed with those from the SM $W^\pm$ boson contributions from $\mathcal L^W$ given by~(\ref{LW}). 
Note that the mediation of $\mathcal D$, $\mathcal H^\pm$ and $\mathcal W$ are always involved a SM fermion and a new hidden heavy fermion in G2HDM.  
Feynman diagrams contributing to $b \to s \gamma$ from $W^\pm$, $\mathcal D$, $\mathcal H^\pm$ and $\mathcal W^{(p,m)}$ are depicted in Figs.~\ref{fig:bsgammaW}, ~\ref{fig:bsgammaD}, \ref{fig:bsgammaH} and~\ref{fig:bsgammaWprime} respectively. 
Needless to say, for the gluon case $b \to s g$, one simply  replace the photon line attached to the colored quarks in these diagrams by the gluon appropriately and hence we will not bother to depict again here.

\section{Wilson Coefficients for $b \to s (\gamma , g) $\label{sec:WilsonCoeffs}}

As mentioned before, the processes $b \to s (\gamma , g)$ can be described by the following effective Lagrangian
\bea
\label{Leff}
\mathscr{L}_{\rm eff} = & - &\frac{1}{32 \pi^2} e \, m_b \, \overline s \sigma_{\mu\nu}
\left( A^M + i \gamma_5 A^E \right) b \, F^{\mu\nu} \nonumber \\ & + & \frac{1}{32 \pi^2} g_s\, m_b \,  \overline s \sigma_{\mu\nu} T^a
\left( C^M + i \gamma_5 C^E \right) b \, G_a^{\mu\nu} \; ,
\eea
where $A^M (C^M)$ and $A^E (C^E)$ are the transition (chromo)magnetic and (chromo)electric dipole form factors respectively, and $F^{\mu\nu} (G_a^{\mu\nu})$ is the electromagnetic (gluon) field strength. Our task is to compute and evaluate these form factors for the on-shell photon at one-loop from the SM $W$ boson loop as well as the three new contributions in minimal G2HDM, as described in previous Section~\ref{sec:G2HDMInteractions}. 

The computation is similar to the charged lepton flavor violation process 
$l_i \to l_j \gamma$ as was presented in~\cite{Tran:2022cwh}, so we can simply recycle our previous formulas. 
The total contribution for $b \to s \gamma$ in minimal G2HDM is given by
\beq
A^{(M,E)} = A^{(M,E)} ( W) + \Delta A^{(M,E)} \; ,
\eeq
where
\bea
\label{MEFFTotalPhoton}
\Delta A^{(M,E)} = A^{(M,E)} (\mathcal D) + \left( A_1^{(M,E)} (\mathcal H) + A_2^{(M,E)} (\mathcal H) \right) + A^{(M,E)} (\mathcal W) \; .
\eea
The SM contributions $A^{(M,E)}(W)$ and new contributions $A^{(M,E)}(\tilde D)$, $A_{1,2}^{(M,E)}(\mathcal H)$, and $A^{(M,E)}(\mathcal W)$ can be found in Appendix~\ref{app:LoopAmps}.

Similarly the total contribution to $b \to s g$ is given by
\beq
C^{(M,E)} = C^{(M,E)} ( W) + \Delta C^{(M,E)} \; ,
\eeq
where
\bea
\label{MEFFTotalGluon}
\Delta C^{(M,E)} =  C^{(M,E)} (\mathcal D) + C^{(M,E)} (\mathcal H) + C^{(M,E)} (\mathcal W) \; ,
\eea
with 
\bea
\label{AMEWgluon}
C^{(M,E)} ( W ) &  =  & A^{(M,E)}_2 ( W ) / Q_u \; , \\
C^{(M,E)} (\mathcal D) & =  & A^{(M,E)} (\mathcal D) / Q_d \; , \\
C^{(M,E)} (\mathcal H) &  =  & A^{(M,E)}_2 (\mathcal H) / Q_u \; , \\
C^{(M,E)} (\mathcal W) &  =  & A^{(M,E)} (\mathcal W) / Q_d \; .
\eea 

In the $B$-physics community, the processes $b \to s (\gamma , g)$ are usually described by the effective Hamiltonian as~\cite{Wen:2023pfq} 
\bea
\label{Heff}
\mathscr{H}_{\rm eff} & = & - \frac{4G_F}{\sqrt 2} \frac{e^2}{16 \pi^2} V_{tb} V_{ts}^* \left[  
 C_7 (\mu) \mathcal O_7 (\mu)  + C_7^\prime (\mu) \mathcal O_7^\prime (\mu)  \right. \nonumber \\
 && \quad\quad\quad\;\;\;\;\;\;\;\;\;\;\;\;\; \left. + \;  C_8 (\mu) \mathcal O_8 (\mu) + C_8^\prime (\mu) \mathcal O_8^\prime (\mu) \right] \,  + \rm{h.c.} \; , 
\eea
with the  operators
\bea
\label{O7O7prime}
\mathcal O_7 & = & \frac{m_b}{e} \overline s \sigma_{\mu\nu}  P_R b F^{\mu\nu} \; , \;\;\;\; \mathcal O_7^\prime = \frac{m_b}{e} \overline s \sigma_{\mu\nu}  P_L b  F^{\mu\nu} \; , \\
\label{O8O8prime}
\mathcal O_8 & = & g_s \frac{m_b}{e^2} \overline s \sigma_{\mu\nu} T^a P_R b G^{\mu\nu}_a \; , \;\;\;\; \mathcal O_8^\prime = g_s \frac{m_b}{e^2} \overline s \sigma_{\mu\nu} T^a P_L b  G^{\mu\nu}_a \; ,
\eea
where $P_{L,R} = (1 \mp \gamma_5)/2$ are the chiral projection operators, $G_F$ is the Fermi constant and 
$C_{7,8}^{(\prime)}(\mu)$ are the dimensionless Wilson coefficients at the scale $\mu$. These Wilson coefficients contain two parts:
\beq
C_{7,8}^{(\prime)}(\mu) = C_{7,8 \; \rm{SM}}^{(\prime)}(\mu) + \Delta C_{7,8}^{(\prime)}(\mu) \; .
\eeq
Comparing the effective Hamiltonian $\mathscr H_{\rm eff}$ in (\ref{Heff}) with $(-\mathscr L_{\rm eff})$ in 
(\ref{Leff}), we can read off the Wilson coefficients  $C_{7,8}^{(\prime)}(M)$ at a high mass scale $M$ where the heavy particles in G2HDM are integrated out. 
Explicitly, we found for the SM $W$ contribution
\bea
\label{C7C7primeSM}
\left(  C_{7 \; \rm{SM}} (M)  ,   C^\prime_{7 \; {\rm SM}} (M) \right) 
& = & - \left( \frac{8 G_F}{\sqrt 2} \right)^{-1} \left( V_{tb} V_{ts}^* \right)^{-1} \nonumber \\
&& \times \left( A^M (W) + i A^E (W) , A^M (W)- i A^E (W) \right) \; ,
\eea
\bea
\label{C8C8primeSM}
\left(  C_{8 \; \rm{SM}} (M)  ,   C^\prime_{8 \; {\rm SM}} (M) \right) 
& = & + \left( \frac{8 G_F}{\sqrt 2} \right)^{-1} \left( V_{tb} V_{ts}^* \right)^{-1} \nonumber \\ 
&& \times \left( C^M (W)  + i C^E (W) , C^M (W) - i C^E (W) \right) \; , 
\eea
and the new contributions from G2HDM
\bea
\left( \Delta C_7(M)  ,  \Delta C^\prime_7(M) \right) 
& = & - \left( \frac{8 G_F}{\sqrt 2} \right)^{-1} \left( V_{tb} V_{ts}^* \right)^{-1} \nonumber \\
&& \times \left( \Delta A^M + i \Delta A^E , \Delta A^M - i \Delta A^E \right) \; , 
\eea
\bea
\left( \Delta C_8(M), \Delta C^\prime_8(M) \right) 
& = & + \left( \frac{8 G_F}{\sqrt 2} \right)^{-1} \left( V_{tb} V_{ts}^* \right)^{-1} \nonumber \\
&& \times \left( \Delta C^M + i \Delta C^E , \Delta C^M - i \Delta C^E \right) \; ,
\eea
where $\Delta A^{(M,E)}$ and $\Delta C^{(M,E)}$ are defined in (\ref{MEFFTotalPhoton}) and (\ref{MEFFTotalGluon}) respectively.

One can then use QCD renormalization group equations to evolve the Wilson coefficients down to $\mu = m_b$ for evaluation of the hadronic matrix elements for $b \to s$ transitions.

\section{Flavor Phenomenology}
\label{sec:fpheno}
Both inclusive and exclusive decays will be taken into account in the following analysis. 
For the inclusive decay, 
{the branching fraction is 
\begin{equation}
\begin{aligned}
	\mathcal{B}\left(B_q\to X_s\gamma\right)\!&=\!\mathcal{B}\left(\bar{B}\to X_c e\bar{\nu}\right)_{\text{exp}}\Big|\frac{V_{ts}^\ast V_{tb}}{V_{cb}}\Big|^2\frac{6\alpha_e}{\pi C}\Delta(E_0;C^{(\prime)\text{eff}}_{7,8}) \, ,\\
	\Delta(E_0;C^{(\prime)}_{7,8})\!&=\!\left[P(E_0;C^{(\prime)\text{eff}}_{7,8}) + N(E_0;C^{(\prime)\text{eff}}_{7,8}) +\epsilon_{\text{EM}}(C^{(\prime)\text{eff}}_{7,8})\right] \, ,
\end{aligned}
\end{equation} 
 where $ C,~P(E_0),~N(E_0) $ and $ \epsilon_{\text{EM}} $ are functions of $C_{7,8}^{(\prime){\rm eff}}$ and their expressions are given in~\cite{Gambino:2013rza,Misiak:2006ab,Gambino:2001ew,Belanger:2004yn}. 
 }
The unprimed effective coefficients can be evaluated via the RGE evolution	
\begin{equation}
	\mu\frac{d\vec{C}^{\text{eff}}(\mu)}{d\mu}=\gamma^T\vec{C}^\text{eff}(\mu)\, ,
\end{equation}	
where the anomalous dimension matrix is defined as $\gamma=\sum\limits_{n=0}\gamma^{n}\left(\frac{\alpha_s(\mu)}{4\pi}\right)^{n+1}$ 
with $\gamma^n$ of NLO \cite{Chetyrkin:1996vx,Gambino:2003zm} and NNLO \cite{Bobeth:2003at,Huber:2005ig,Czakon:2006ss}.
For the effective coefficients  \cite{Buras:1993xp,Chetyrkin:1996vx,Greub:1996tg} at EW scale, here we adopt the convention $C_{7,8}^{\text{eff}}=C_{7,8}+\sum\limits_{j=1}^6 y^{(7,8)}_j C_j$
with $y_j^{(7)}=\left(0,0,-\frac13,-\frac49,-\frac{20}{3},-\frac{80}{9} \right)$ and 
$y_j^{(8)}=\left(0,0,1,-\frac16,20,-\frac{10}{3} \right)$. 
Notice in the practical calculation we have neglected new physics contributions to four-quark operators $O_{1,\ldots, 6}$.
Since new particles in G2HDM are supposed not to emerge between $\mu_W $ and $ \mu_{t} $,
the primed coefficients $ C_{7,8}^{\prime \, \rm{eff}}$ share the common evolution equations with their chiral-flipped counterparts \cite{Everett:2001yy,Eberl:2021ulg, Borzumati:1999qt}.

The branching fraction of exclusive radiative decay $ B\to V\gamma $  can be generally written \cite{Paul:2016urs} as
\begin{equation}
	\begin{aligned}
		\mathcal{B}\left(B_q\to V\gamma\right)=\tau_{B_q}\frac{\alpha_eG_F^2m^3_{B_q}m_b^2}{32\pi^4}\left(1-\frac{m_V^2}{m_B^2}\right)^3|\lambda_t|^2\left(|C_7^{\text{eff}}|^2+|C_7^{\prime\text{eff}}|^2\right)T_1(0) \, ,
	\end{aligned}
\end{equation} 
where the final state dependent form factors $T_1(0)$ are taken from \cite{Paul:2016urs}
based on a combination with light-cone sum rule and Lattice QCD.
	
In the normalized CP asymmetry for $B_s\to V\gamma$, assuming its parametrization obeying generic time dependent form
\footnote{In the parametrization $\mathcal{A}_{\text{CP}}(B_s\to V\gamma)[t]=\frac{S \sin(\Delta m_s t)- C \cos(\Delta m_s t)}
{\cosh(\frac12 \Delta \Gamma_s t)-H \cosh(\frac12 \Delta \Gamma_s t)}$,
we adopt the convention $C_{V\gamma}=C$, $S_{V\gamma}=S$, $A^{\Delta}_{V\gamma}=H$ in this work.
}, the observables are defined  \cite{Muheim:2008vu}  as 	
\begin{equation}
	\begin{aligned}
		C_{V\gamma}&=\frac{|\mathcal{A}_L|^2+|\mathcal{A}_R|^2-|\bar{\mathcal{A}}_R|^2-|\bar{\mathcal{A}}_L|^2}{|\mathcal{A}_L|^2+|\bar{\mathcal{A}}_L|^2+|\mathcal{A}_R|^2+|\bar{\mathcal{A}}_R|^2} \; , \\
		S_{V\gamma}&=2 \Im\left[\frac{\frac{q}{p}(\bar{\mathcal{A}}_L\mathcal{A}_L^\ast+\bar{\mathcal{A}}_R\mathcal{A}_R^\ast)}{|\mathcal{A}_L|^2+|\bar{\mathcal{A}}_L|^2+|\mathcal{A}_R|^2+|\bar{\mathcal{A}}_R|^2}\right] \, , \\
		A^{\Delta}_{V\gamma}&=2 \Re\left[\frac{\frac{q}{p}(\bar{\mathcal{A}}_L\mathcal{A}_L^\ast+\bar{\mathcal{A}}_R\mathcal{A}_R^\ast)}{|\mathcal{A}_L|^2+|\bar{\mathcal{A}}_L|^2+|\mathcal{A}_R|^2+|\bar{\mathcal{A}}_R|^2}\right] \, ,
	\end{aligned}
	\label{eq:obs}
\end{equation}
in terms of the amplitudes
$\mathcal{A}_{L(R)}
=\mathcal{N}C_7^{(\prime)\text{eff}}T_{1}(0)$ and $\bar{\mathcal{A}}_{L(R)} 
 \equiv \mathcal{A}(\bar{B}_s\to V\gamma_{L(R)})$  with $\mathcal{N}=\lambda_t \sqrt{\frac{G_F^2\alpha_em_B^3}{32\pi^4}\left(1-\frac{m_V^2}{m_B^2}\right)^3}$. 
 In particular, $\bar{\mathcal{A}}_{L(R)}$ can be derived straightforwardly from ${\mathcal{A}}_{L(R)}$
 by taking weak phase conjugated while keeping strong phase unchanged.
The defined ratio is $\left( \frac{q}{p} \right)_s=\left|  \frac{q}{p}\right|_s e^{-i\phi_s}$ and $\left|  \frac{q}{p}\right|_s=1$ has been utilized
to derive Eq. (\ref{eq:obs}).
\footnote{
{Here we simply adopt the experimental average value
$\phi_s=-0.008\pm0.019$
 \cite{HFLAV:2022esi} in the following numerical analysis.
 }}

To date, the branching fractions of $B^{(0,+)}\to K^{\ast (0,+)} \gamma$, $B\to \phi\gamma$ 
and CP asymmetry parameters of  $B_s\to \phi \gamma$ have been measured, which can be taken
as input in the following numerical analysis.

\section{Numerical Analysis\label{sec:Numerics}}
In this section, we present the numerical results including the new contributions to the Wilson Coefficients  $C_7^{(\prime)}$ and $C_8^{(\prime)}$ and the preferred regions on the model parameter space for data from various low-energy flavor observables as well as the constraints derived from theoretical conditions on the scalar potential and oblique parameters. In this analysis, we assume the new mixing matrices $V_u^H \equiv V_d^H \equiv V_{\rm CKM}$ and fix the hidden quark masses as $m_{s^H} = m_{d^H}$, $m_{b^H} = m_{d^H} + \Delta m_{d^H}$ for down-type quarks, and $m_{c^H} = m_{u^H}$, $m_{t^H} = m_{u^H} + \Delta m_{u^H}$ for up-type quarks.

\begin{figure}[htbp]
    \centering
    \includegraphics[width=0.45\linewidth]{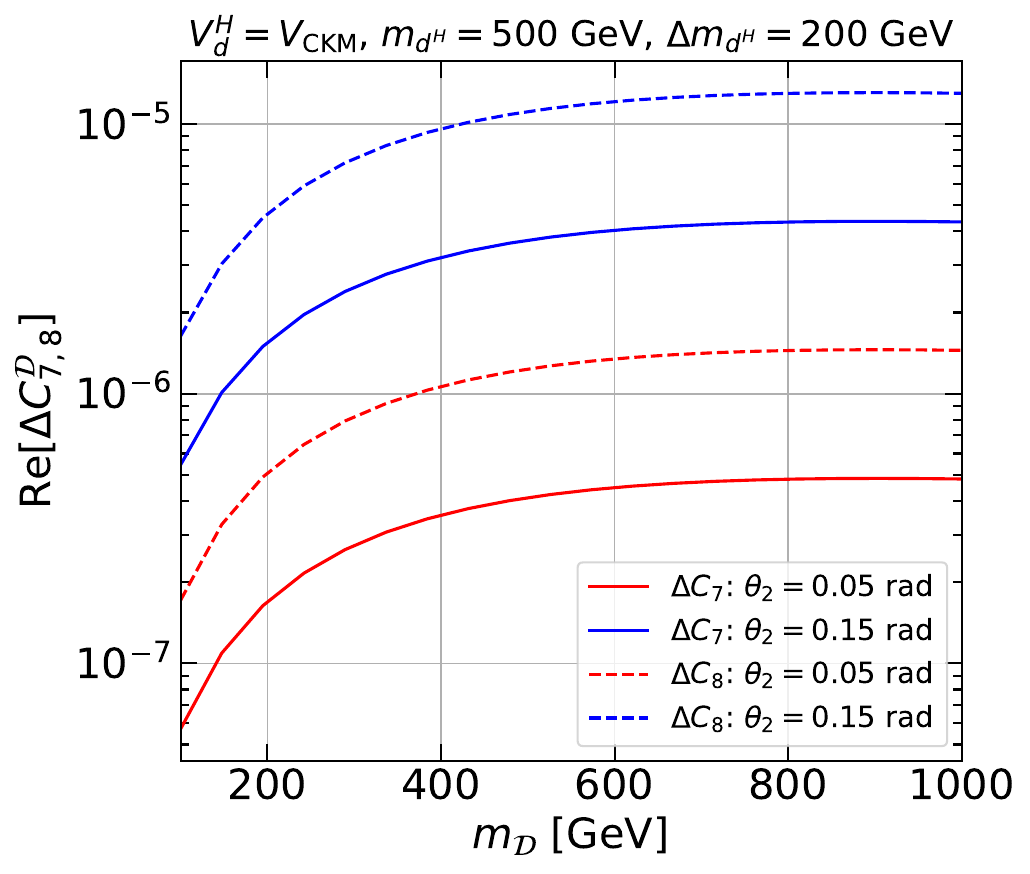}
    \includegraphics[width=0.45\linewidth]{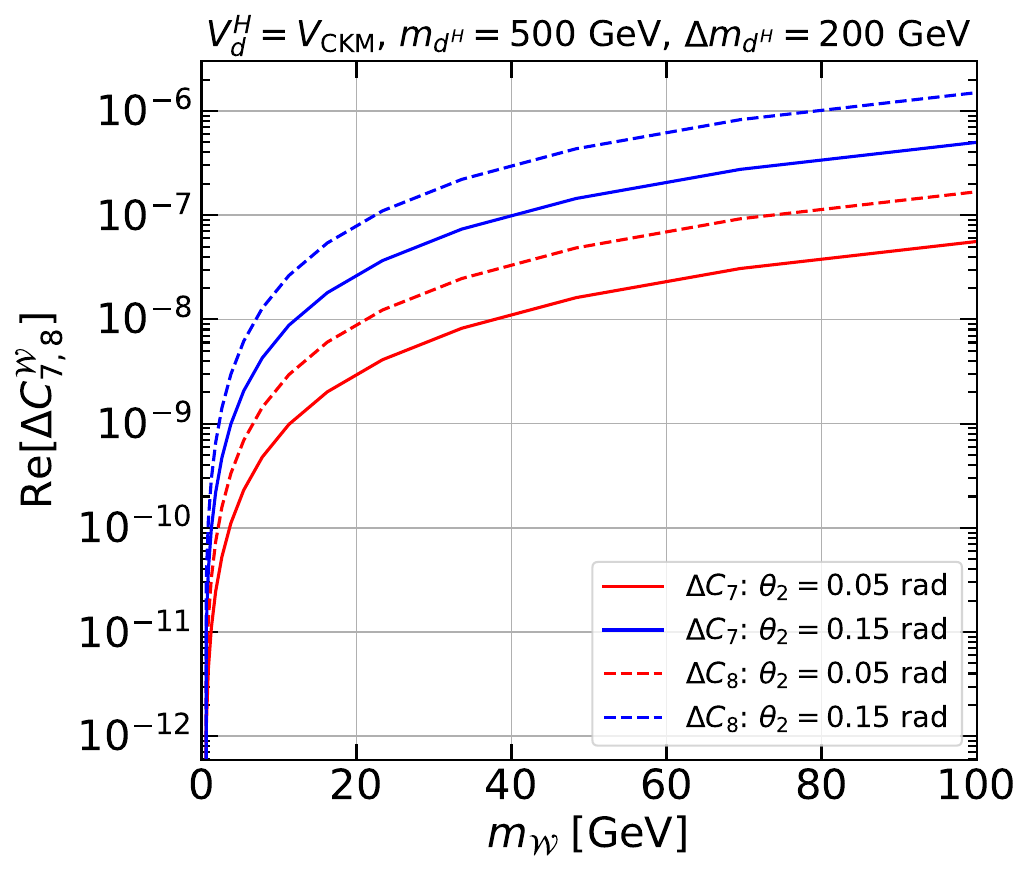}
    \includegraphics[width=0.45\linewidth]{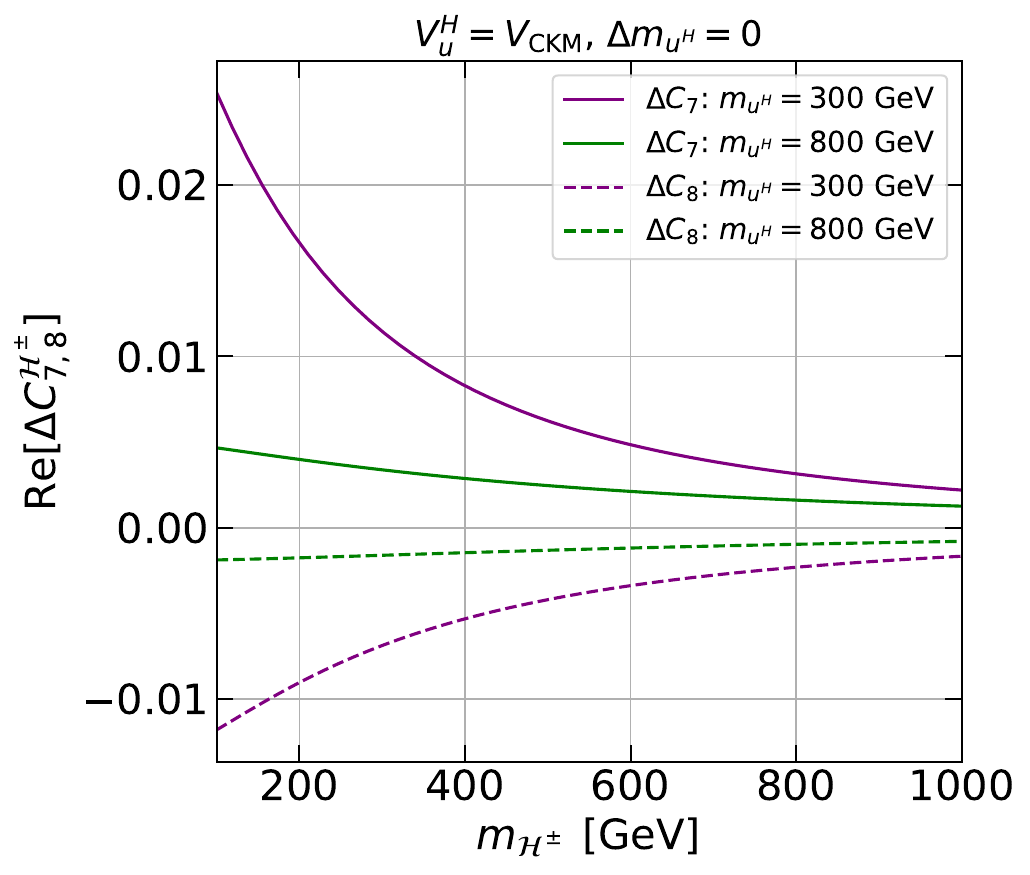}
    \includegraphics[width=0.45\linewidth]{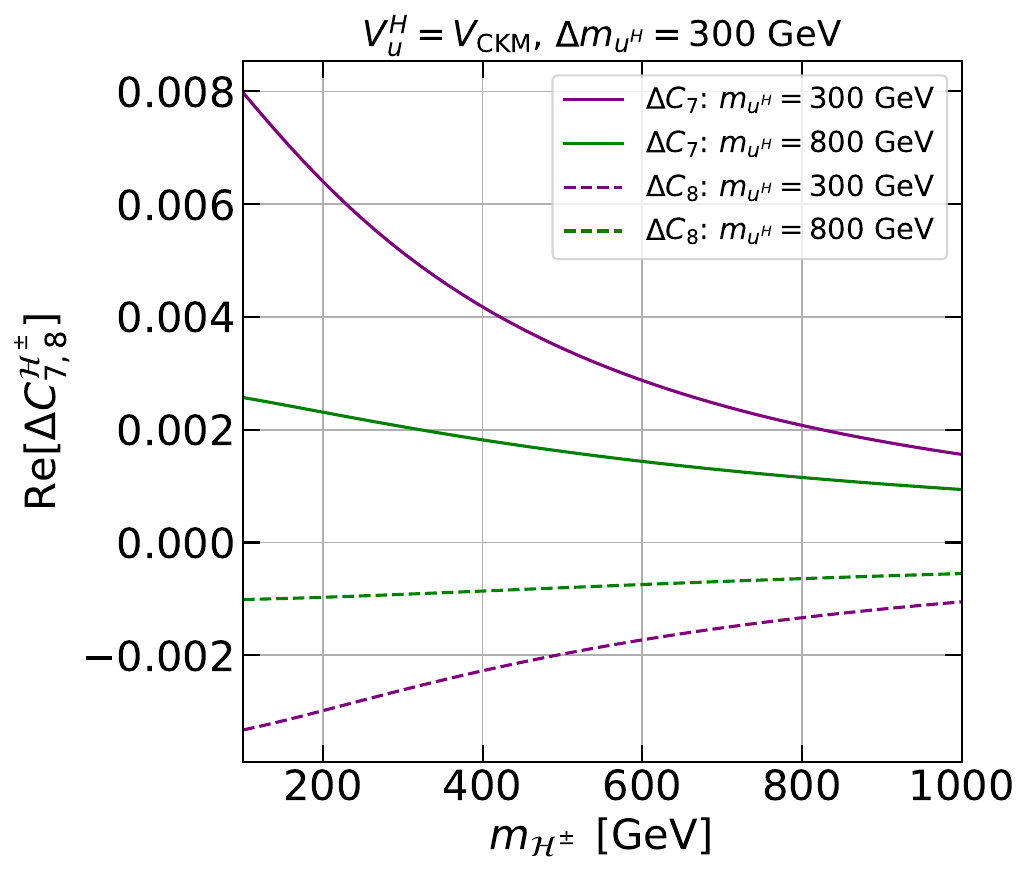}
    \caption{Real part of the new contributions to the Wilson coefficients $\Delta C_7$ and $\Delta C_8$. 
    Top panels: Contribution from the $\mathcal{D}$ loop as a function of $m_{\mathcal{D}}$ (left) and from the $\mathcal{W}$ loop as a function of $m_{\mathcal{W}}$ (right). We set $V_d^H = V_{\rm CKM}$, $m_{d^H} = 500$ GeV, and $\Delta m_{d^H} = 200$ GeV. The solid (dashed) blue and red lines indicate Re[$\Delta C_7$] (Re[$\Delta C_8$]) with fixed $\theta_2 = 0.05$ rad and $\theta_2 = 0.15$ rad, respectively.
    Bottom panels: Contribution from the ${\mathcal H}^{\pm}$ loop as a function of $m_{\mathcal{H}^\pm}$, with $V_u^H = V_{\rm CKM}$ and $\Delta m_{u^H} = 0$ GeV for the left panel, and $\Delta m_{u^H} = 300$ GeV for the right panel. The solid (dashed) purple and green lines represent Re[$\Delta C_7$] (Re[$\Delta C_8$]) with fixed $m_{u^H} = 300$ GeV and $m_{u^H} = 800$ GeV, respectively.
    }
    \label{fig:DeltaC7C8}
\end{figure}

\begin{figure}[htbp]
    \centering
    \includegraphics[width=0.45\linewidth]{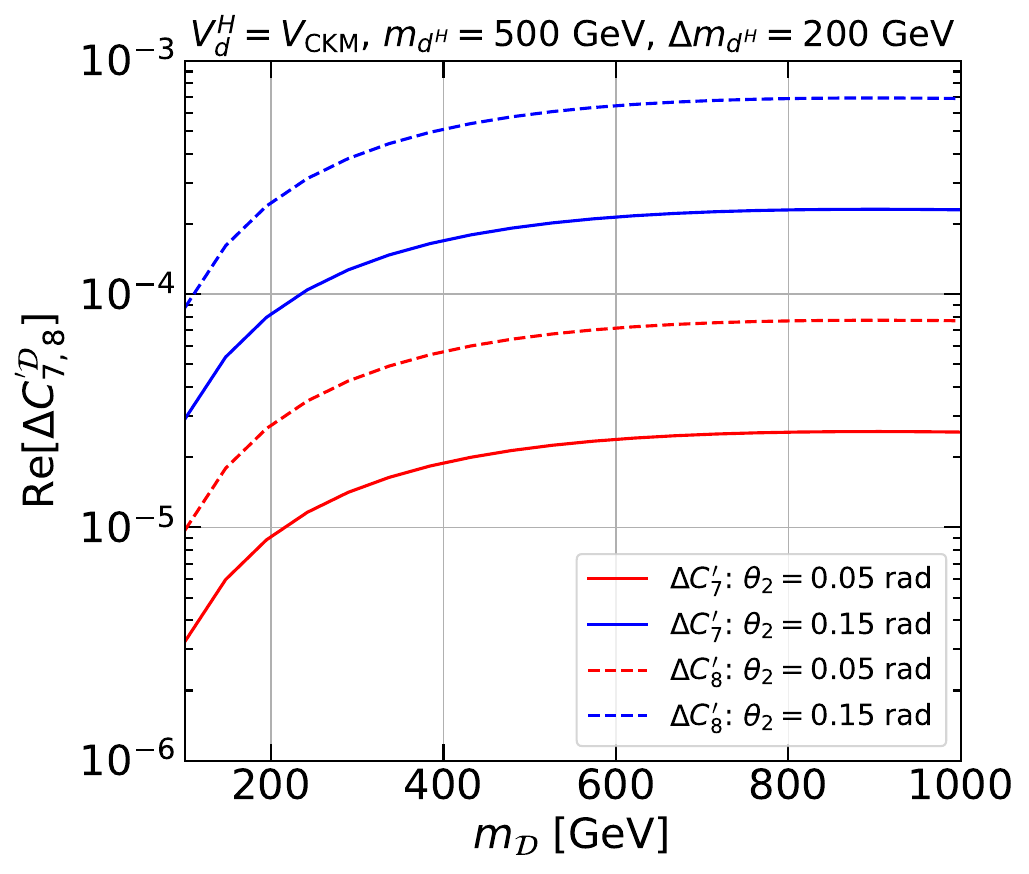}
    \includegraphics[width=0.45\linewidth]{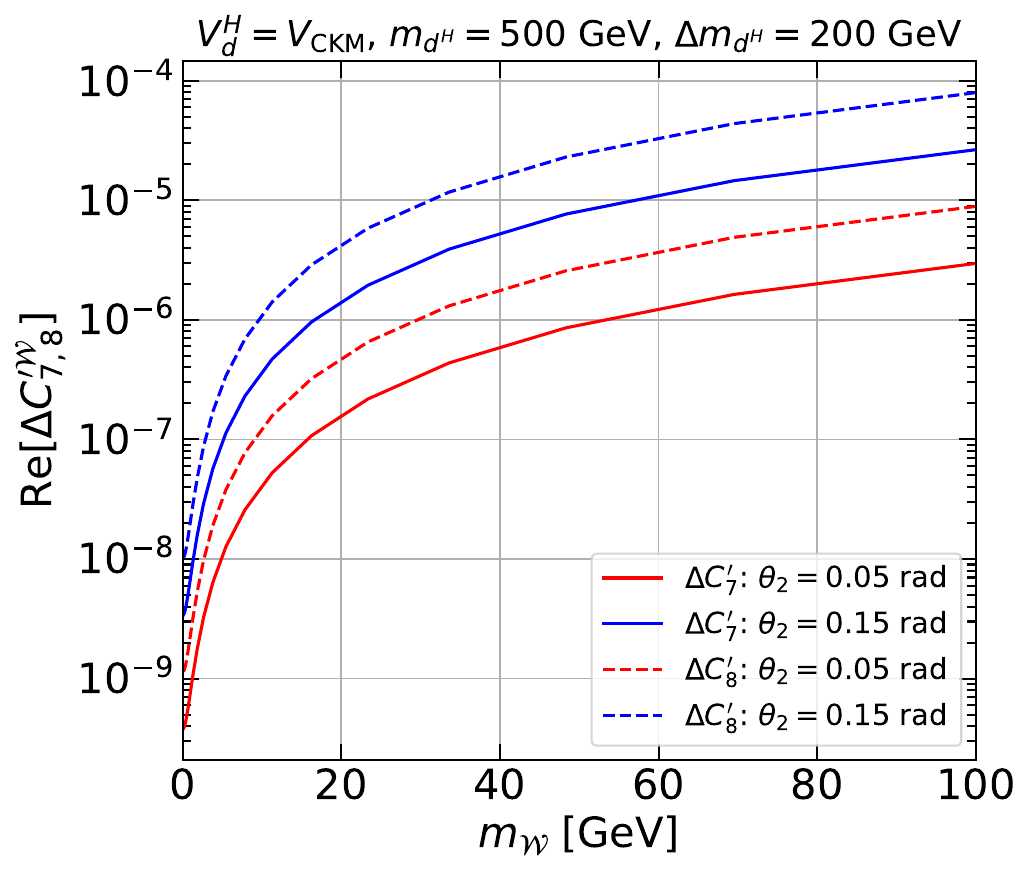}
    \includegraphics[width=0.45\linewidth]{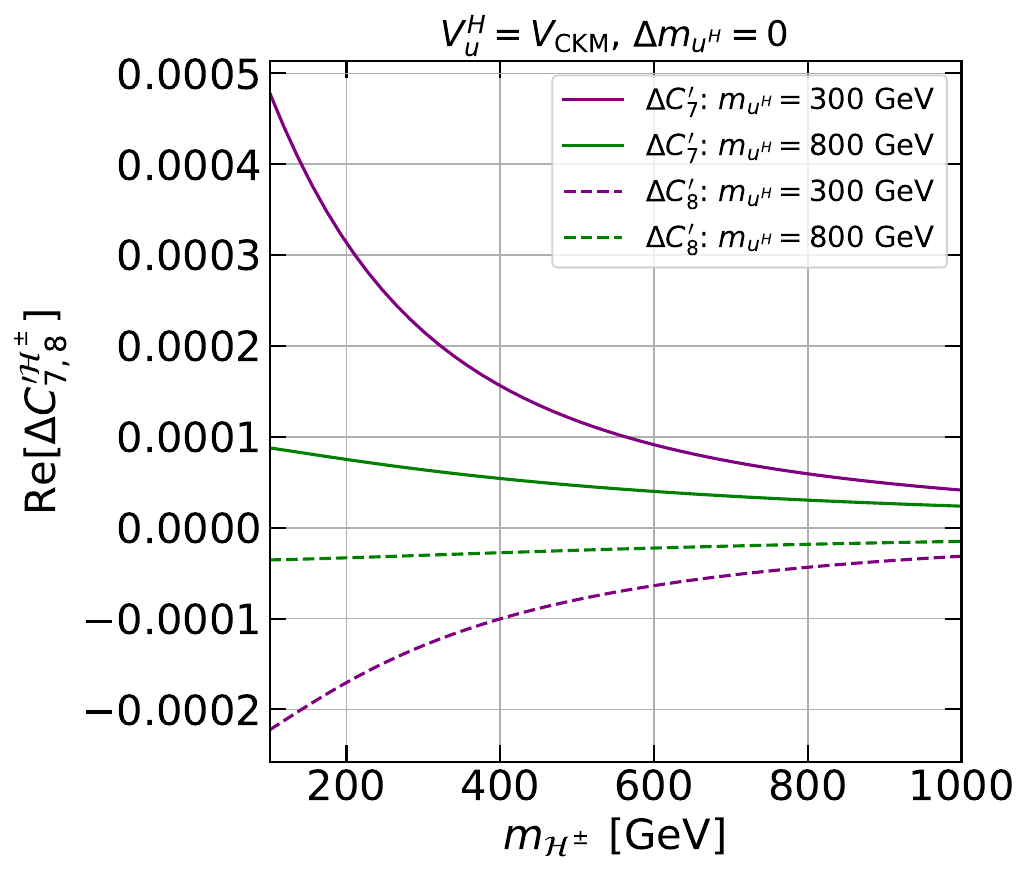}
    \includegraphics[width=0.45\linewidth]{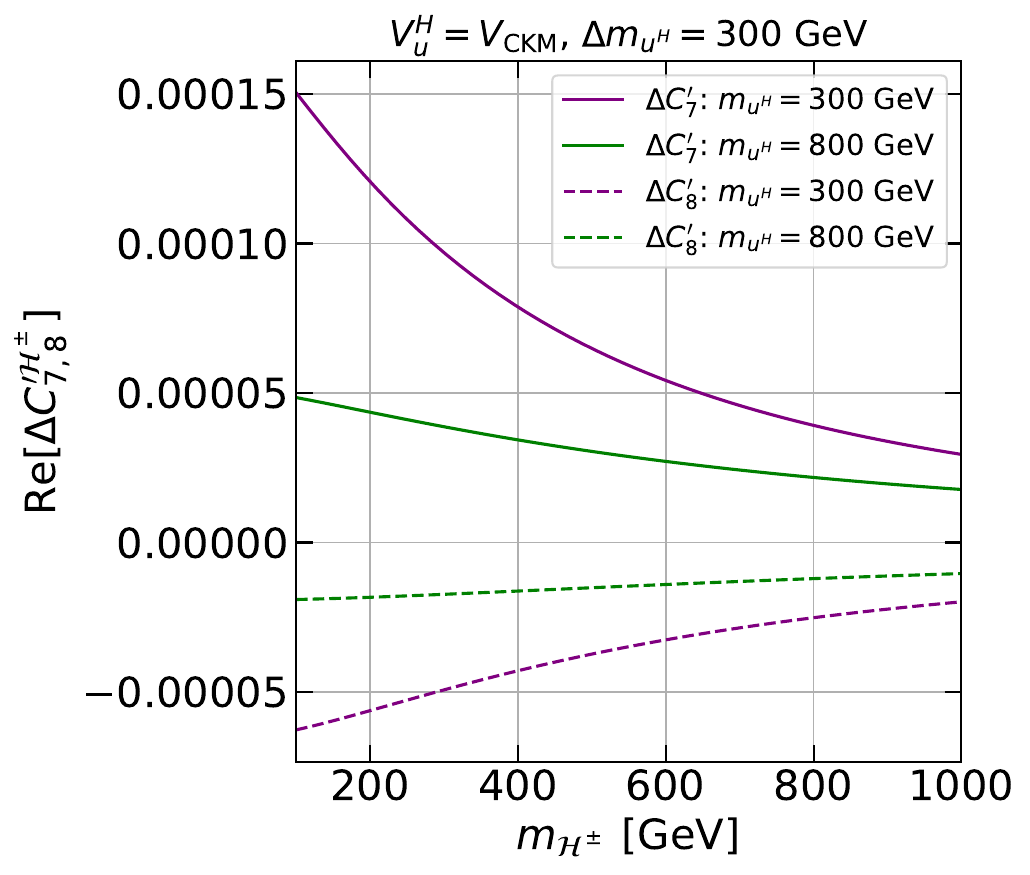}
    \caption{Similar to Fig.~\ref{fig:DeltaC7C8} but for $\Delta C_7'$ and $\Delta C_8'$. 
    }
    \label{fig:DeltaC7C8p}
\end{figure}

\subsection{New Contributions to the Wilson Coefficients  $C^{(\prime)}_7$ and $C_8^{(\prime)}$}

Before delving into an examination of the parameter space within the model in light of observations from low-energy flavor experiments, we would like to know what are the relative magnitudes of the new contributions to the Wilson coefficients $\Delta C_7^{(\prime)}$ and $\Delta C_8^{(\prime)}$.

Fig.~\ref{fig:DeltaC7C8} illustrates the real part of $\Delta C_7$ and $\Delta C_8$~\footnote{The imaginary parts of $\Delta C_7^{(\prime)}$ and $\Delta C_8^{(\prime)}$ are significantly smaller, typically by at least two orders of magnitude, compared to their real parts.}. The top panels present the contributions from $\mathcal{D}$ loop and $\mathcal{W}$ loop diagrams. We illustrate the results with fixed values of $\theta_2 = 0.05$ rad (red lines) and $\theta_2 = 0.15$ rad (blue lines)\footnote{The choice of $\theta_2$ values adheres to the current constraints for the light mass DM candidate ($\theta_2 \leq 0.15$ rad), as investigated in Ref.~\cite{Tran:2022yrh}.}.

We find that $\Delta C_7$ and $\Delta C_8$ are highly dependent on the values of the mixing angle $\theta_2$ and the masses involved in the loop.
A smaller $\theta_2$ and lower $m_{\mathcal{D}}$ or $m_{\mathcal{W}}$ result in diminished contributions to both Re[$\Delta C_7$] and Re[$\Delta C_8$]. Here, we set $V_d^H = V_{\rm CKM}$, $m_{d^H} = 500$ GeV, and $\Delta m_{d^H} = 200$ GeV. It's worth noting that if $\Delta m_{d^H} = 0$, indicating degenerate masses of heavy down-type quarks, both the contributions from $\mathcal{D}$ loop and $\mathcal{W}$ loop diagrams vanish. 

Conversely, the contribution from ${\mathcal H}^{\pm}$ loop diagrams doesn't vanish when the masses of heavy up-type quarks are degenerate ({\it i.e.}, $\Delta m_{u^H} = 0$), but it is enhanced compared to the non-degenerate case, as depicted in the bottom panels of Fig.~\ref{fig:DeltaC7C8}. 
Furthermore, the contribution is more significant in regions of lighter $m_{{\mathcal H}^\pm}$ and $m_{u^H}$. Unlike the $\mathcal{D}$ loop and $\mathcal{W}$ loop diagrams, which provide positive contributions to both Re[$\Delta C_7$] and Re[$\Delta C_8$], ${\mathcal H}^{\pm}$ loop diagrams yield a positive value for Re[$\Delta C_7$] and a negative value for Re[$\Delta C_8$] in the parameter space of interest. Ultimately, we find that the contribution from ${\mathcal H}^{\pm}$ loop diagrams dominates over the $\mathcal{D}$ loop and $\mathcal{W}$ loop diagrams.
This implies that the relic density constraint from the dark matter candidate $\mathcal{W}$ is not relevant in our analysis.

We note that if one also takes the up-type SM quark masses (in additional to degenerate up-type hidden heavy quark masses) to be the same, the charged Higgs $\mathcal{H^\pm}$ loop diagram also vanish too, just like the SM $W^\pm$ loop. All these null results for the $W^\pm$, $\mathcal{W}$, $\mathcal{D}$ and $\mathcal{H}^\pm$ loop contributions in the degenerate mass scenarios are just manifestation of a generalized version of GIM mechanism~\cite{Glashow:1970gm} in G2HDM.

Figure \ref{fig:DeltaC7C8p} presents the real part of $\Delta C_7^\prime$ and $\Delta C_8^\prime$, with the parameter space setup identical to that of Figure \ref{fig:DeltaC7C8}. We observe a similar dependence of $\Delta C_7^\prime$ and $\Delta C_8^\prime$ on the relevant parameter space as seen from $\Delta C_7$ and $\Delta C_8$.
Moreover, for the $\mathcal{D}$ loop and $\mathcal{W}$ loop diagrams, $\Delta C_7^\prime$ ($\Delta C_8^\prime$) can be approximately two orders of magnitude larger than \(\Delta C_7\) (\(\Delta C_8\)) within the same parameter space of interest. This discrepancy primarily arises from the presence of right-handed currents in the $\mathcal{D}$ loop and $\mathcal{W}$ loop diagrams.
On the other hand, the contribution from \(\mathcal{H}^\pm\) loop diagrams to $\Delta C_7^\prime$ ($\Delta C_8^\prime$) is approximately two orders of magnitude smaller compared to its contribution to $\Delta C_7$ ($\Delta C_8$).

\subsection{Analysis Strategy and Inputs}

In Bayesian analysis,  
the posterior function is proportional to the product of likelihood function and prior, giving
\begin{equation}
	\begin{aligned}
		\mathcal{P}(\vec{\theta}|\mathcal{O}_{\text{expt.}})\propto \mathcal{L}(\mathcal{O}|\vec{\theta})\pi(\vec{\theta}) \, .
	\end{aligned}
\label{eq:posterior}
\end{equation}
The Negative Log Likelihood (NLL) function is further defined via $\chi^2$ as
\begin{equation}
	\begin{aligned}
		-2\log\mathcal{L}(\mathcal{O}|\vec{\theta})&=\chi^2(\vec{\theta})\\
		&=(\mathcal{O}_{\text{theo.}}(\vec{\theta})-\mathcal{O}_{\text{expt.}})^\top(V_{\text{expt.}}+V_{\text{theo.}})^{-1}(\mathcal{O}_{\text{theo.}}(\vec{\theta})-\mathcal{O}_{\text{expt.}}) \, ,
	\end{aligned}
\end{equation}
where $ \mathcal{O}_{\text{theo.}} $ and $ \mathcal{O}_{\text{expt.}} $ denote
the theoretical predictions and experimental values, respectively,
 of observables of interest. 
 Additionally, 
 the covariance matrices $ V_{\text{theo.}} $ and $ V_{\text{expt.}} $ incorporate 
 their respective errors. 
The set of observables, including  pseudo-observables 
 $ C_{8}^{(\prime)}(\mu_b)$ \footnote{
Although they are not actual observables, the information extracted from the model-independent global fit~\cite{Wen:2023pfq} is essential for imposing additional constraints on a detailed model.}, are summarized as 
 \begin{equation}
	\mathcal{O}^\top=\begin{aligned}
		\bigl[\mathcal{B}(B\to\phi\gamma),~\bigr.
		\mathcal{B}(B\to K^{\ast0}\gamma),~
		&\mathcal{B}(B\to K^{\ast+}\gamma),~
		\mathcal{B}(B\to X_s\gamma),\\
		S_{K^\ast\gamma},
		S_{\phi\gamma},
		C_{\phi\gamma},
		&A^\Delta_{\phi\gamma},
		C_{8},
		\bigl.C^\prime_{8}\bigr] \, ,
	\end{aligned}
 \label{observables}
\end{equation}
with corresponding experimental values collected in Table \ref{tab:set1}. For our keen readers, other inputs entered implicitly in the eight observables in (\ref{observables}) for carrying out the theoretical calculations are summarized
in Table \ref{tab:Input_para} as well.

\begin{table}[t]
	\caption{Experimental values of observables 
	related to $ C_{7}/C_{7}' $.} \label{tab:set1}
	\resizebox{\textwidth}{!}{
	\begin{tabular}{ccccc}
		\hline\hline
		\makebox[0.12\textwidth][c]{}&\makebox[0.27\textwidth][c]{$10^5\mathcal{B}$}&\makebox[0.27\textwidth][c]{$ S $}& \makebox[0.27\textwidth][c]{$ C $} &\makebox[0.27\textwidth][c]{$ A^{\Delta} $}\\
		\hline
		$B\to X_s\gamma$ &$37.5\pm1.8\pm3.5$\cite{Belle:2014nmp}&-&-&-\\
		$B\to \phi\gamma$&$3.6\pm0.5\pm0.3\pm0.6$\cite{Belle:2014sac}&$0.43\pm0.30\pm0.11$\cite{LHCb:2019vks}&$0.11\pm0.29\pm0.11$\cite{LHCb:2019vks}&$-0.67^{+0.37}_{-0.41}\pm0.17$\cite{LHCb:2019vks}\\
		$B\to K^{\ast0}\gamma$&$4.5\pm0.3\pm0.2$\cite{BelleII:2021tzi}&$-0.16\pm0.22$\cite{HFLAV:2022esi}&-&-\\
		$B\to K^{\ast+}\gamma$&$5.2\pm0.4\pm0.3$\cite{BelleII:2021tzi}&-&-&-\\
		\hline
		\hline
	\end{tabular}}
\end{table}

\begin{table}[t]
	\centering
	\caption{\label{tab:Input_para}
		Input parameters adopted in theoretical calculations of observables of interest. 	}
	\renewcommand\arraystretch{1.1}
	\begin{tabular}{cccc}
		\hline\hline
		Parameters& Values& Parameters & Values \\
		\hline
		$ \mathcal{Y}_{b}/10^{-2} $& 1.646($8.2$)\cite{Deppisch:2018flu}&$\mathcal{Y}_{t} $ & 0.9897(86)\cite{Deppisch:2018flu} \\
		$ \mathcal{Y}_{c}/10^{-3} $&3.646$(91) $\cite{Deppisch:2018flu}&$ \mathcal{Y}_{s}/10^{-4} $ &3.104$(36)$\cite{Deppisch:2018flu}\\
		$ \mathcal{Y}_{d}/10^{-5} $&1.663$ (64) $\cite{Deppisch:2018flu}&$ \mathcal{Y}_{u}/10^{-6} $&7.80$( 86)$\cite{Deppisch:2018flu}\\
		
		$ m_{B_d} $&5279.65(12)~MeV\cite{PDG2022}&$ m_{B_s} $&5366.92(10)~MeV\cite{PDG2022}\\
		$ m_{B_u} $&5279.34(12)~MeV\cite{PDG2022}&$ m_\phi $&1019.461(16)~MeV\cite{PDG2022}\\
		$ m_{K^{\ast\pm}} $&891.67(26)~MeV\cite{PDG2022}&$ m_{K^{\ast0}} $&895.55(20)~MeV\cite{PDG2022}\\
		$ \tau_{B_s} $&1.520(5)~ps\cite{PDG2022}&$ \tau_{B_u} $&1.638(4)~ps\cite{PDG2022}\\
		
		$ \tau_{B_{d}} $&1.519$\left(4\right)$~ps\cite{PDG2022}
		&$ G_F $&1.1663787(6)~GeV$ ^{-2} $\cite{PDG2022}\\
		\hline
		$\alpha_s(m_Z)$&0.1179(9)\cite{PDG2022}&$\alpha_{\text{e}}(m_{Z})$&1/127.944(14)\cite{PDG2022}\\
		
		$y_s$&0.064(4)\cite{PDG2022}&$y_d$&0.0005(50)\cite{PDG2022}\\
		$ \sin^2\theta_W $&0.23121(4)\cite{PDG2022}&$ \mu^2_{G} $&0.336$(64) $\cite{Gambino:2013rza}\\
		$ \rho_D^3 $&$ 0.153(45) $\cite{Gambino:2013rza}&$ \rho_{LS}^3 $&$ -0.145(98) $\cite{Gambino:2013rza}\\
		$ \mathcal{B}(B\to X_ce\bar{\nu})_{\text{exp}} $&$ 0.0997(41)$\cite{Belle-II:2021jlu}&$ \phi_s $&$ 0.0046(12) $\cite{HFLAV:2022esi}\\
		\hline
		$ \sin\theta_{12} $&0.22500(67)\cite{PDG2022}&$ \sin\theta_{13} $&0.00369$ (11) $\cite{PDG2022}\\
		$\sin\theta_{23}$&$ 0.04182(^{+85}_{-74}) $\cite{PDG2022}&$\delta_{CP}$&$ 1.144(27) $\cite{PDG2022}\\
		\hline\hline
	\end{tabular}
\end{table}

\subsection{Scanning Results}

\begin{figure}[htbp]
	\centering
	\includegraphics[width=0.95\linewidth]{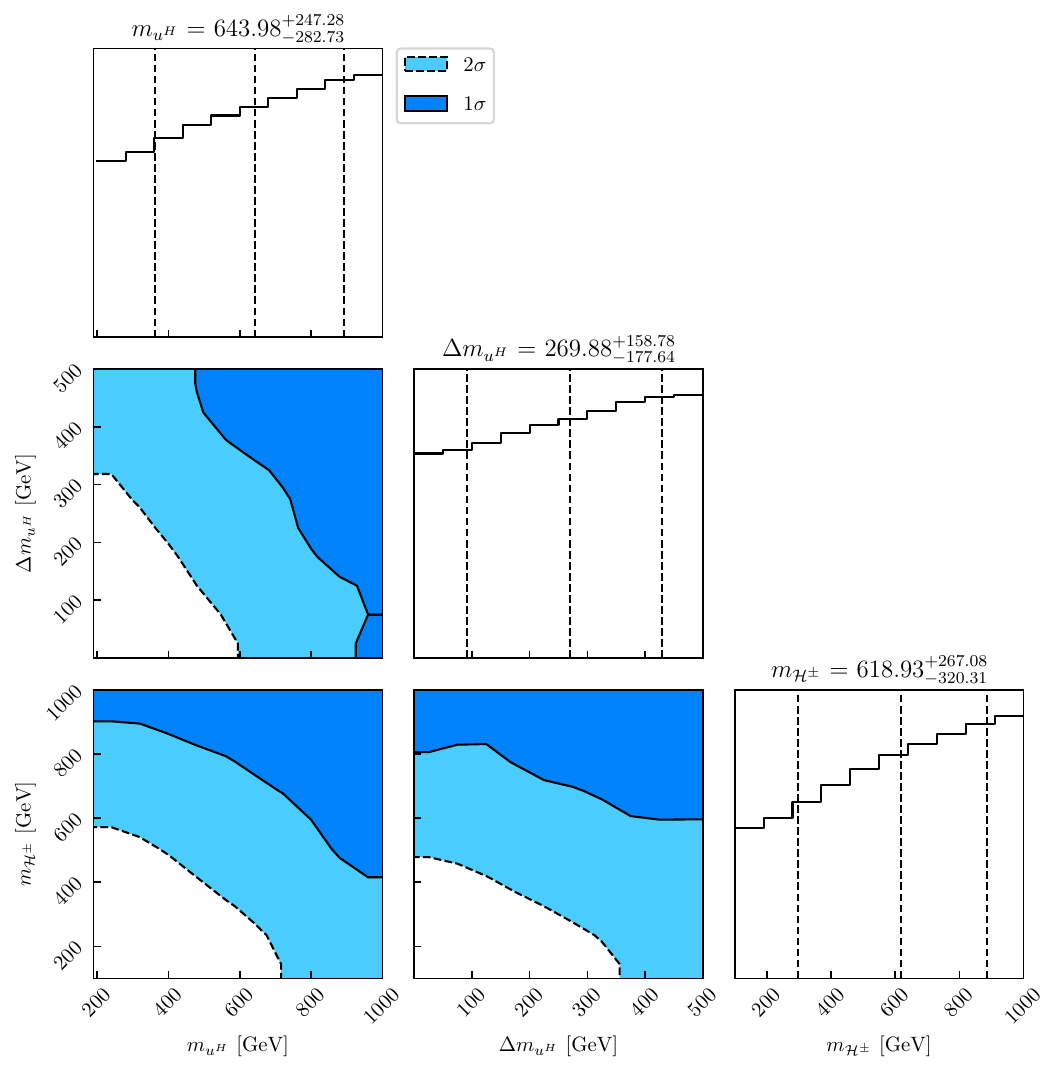}
	\caption{
	The parameter space associated with $ \mathcal{H}^\pm $  constrained by low-energy flavor experiments. The solid (dashed) boundary delineates the region of interest, with darker (lighter) shading indicating areas within a $68\%$ ($95\%$) confidence level.}
	\label{fig:corner1}
\end{figure}

To explore the remaining related parameters, we implement the affine-invariant ensemble sampler for Markov chain Monte Carlo (MCMC)~\texttt{emcee} \cite{Foreman-Mackey:2012any}. Utilizing this method enables the posterior function (\ref{eq:posterior}) to efficiently converge towards solutions with higher probabilities in the parameter space.

We specify the following priors for the remaining parameters in the model:
\begin{equation}
	\begin{aligned}
		\vec{\theta}=\left\{\begin{array}{rcl}
			m_{d^H}\in[200,1000]~\text{GeV},&~
			\Delta m_{d^H}\in[0,500]~\text{GeV} ,\\
			m_{u^H}\in[200,1000]~\text{GeV},&~
			\Delta m_{u^H}\in[0,500]~\text{GeV}  ,\\
			m_{\mathcal{W}}\in[0.01,100]~\text{GeV},&~
			m_{\mathcal{D}}\in[100,1000]~\text{GeV} , \\
			m_{\mathcal{H}^\pm}\in[100,1000]~\text{GeV},&~
			\theta_{2}\in[-\frac{\pi}{2},\frac{\pi}{2}] \, .
		\end{array}\right.
	\end{aligned}
	\label{{eq:prior}}
\end{equation} 
These priors assume that each parameter follows a flat prior probability (uniform distribution), which assists in defining the search intervals.

Fig.~\ref{fig:corner1} illustrates the favored region from low-energy flavor experiments spanned on sensitive parameters. 
We found that the dominant contribution to the $b \to s\gamma$ process arises from charged Higgs diagrams, leading to significant constraints on related parameters such as $m_{\mathcal{H}^\pm}$, $m_{u^H}$, and $\Delta m_{u^H}$, as illustrated in Fig.~\ref{fig:corner1}. 
In particular, one can put a lower bound on the charged Higgs mass depending upon both the hidden up-type quark mass and the mass splitting $\Delta m_{u^H}$. Within the $2\sigma$ confidence interval, we find $m_{\mathcal{H}^{\pm}} \gtrsim 180$ GeV when $m_{u^H} \simeq 700$ GeV and $\Delta m_{u^H} \simeq 350$ GeV. This constraint can become even more stringent in the lower mass range of the hidden up-type quark and for smaller values of the mass splitting.

\begin{figure}[htbp]
	\centering
	\includegraphics[width=0.65\linewidth]{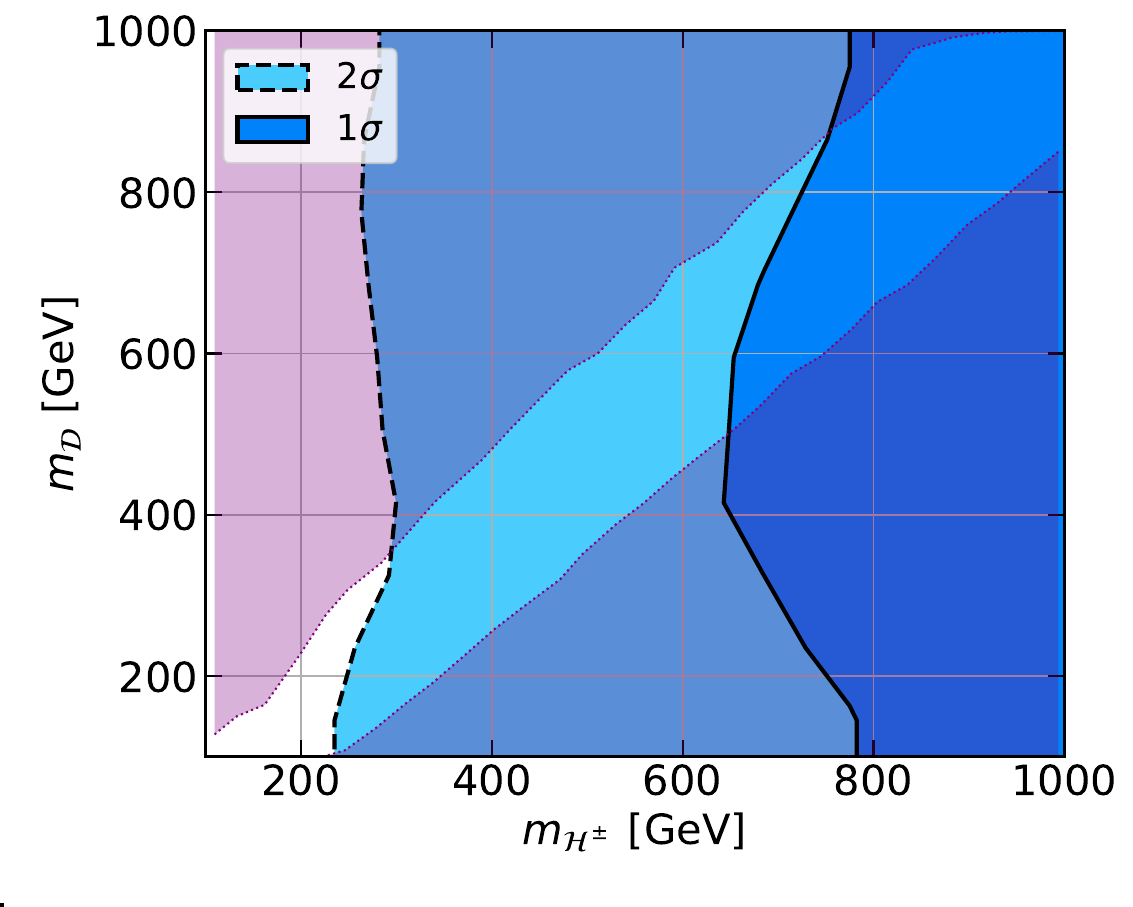}
	\caption{The $68\%$ and $95\%$ confidence level contours from low-energy flavor experiments (lighter blue and darker blue regions), and exclusion regions (purple shaded regions) from a combination of theoretical and oblique parameters constraints projected on the ($m_{\mathcal{H}^\pm}, m_{\mathcal D}$) plane.}
	\label{fig:mcH_mD}
\end{figure}

In Fig.~\ref{fig:mcH_mD}, we present the favored region delineated by low-energy flavor experiments on the ($m_{\mathcal{H}^\pm}, m_\mathcal D$) plane. Owing to the negligible influence of the $\mathcal{D}$ diagram, the bounds derived from these experiments show minimal dependence on $m_\mathcal D$. Within the same figure, we also show exclusion regions (purple regions) from theoretical constraints of the scalar potential, which includes the criteria for vacuum stability and perturbative unitarity~\cite{Ramos:2021omo,Ramos:2021txu}, alongside constraints~\cite{Tran:2022yrh} from oblique parameters~\cite{Peskin:1991sw}. These constraints introduce a significant correlation between $m_{\mathcal{H}^\pm}$ and $m_\mathcal D$. When these are combined with the low-energy flavor experiment constraints, it becomes possible to establish lower bounds on both $m_{\mathcal{H}^\pm}$ and $m_\mathcal D$. Notably, the regions where $m_{\mathcal{H}^\pm} \lesssim 250$ GeV and $m_\mathcal D \lesssim 100$ GeV are excluded at the $95\%$ confidence level through a synergy of constraints from low-energy flavor experiments, oblique parameters, and theoretical constraints imposed on the scalar potential.

We note that because of the minor contributions from $\mathcal{D}$ and $\mathcal{W}$ diagrams, parameters like $m_{d^H}$, $\Delta m_{d^H}$, and $m_{\mathcal{W}}$ remain relatively unconstrained and are not depicted here.

\section{Conclusions\label{sec:Conclusions}}
In this study, we have performed computations of the one-loop radiative decay processes for the flavor-changing bottom quark transitions $b \to s \gamma$ and $b \to s g$ within the framework of a minimal G2HDM. Our analysis extends beyond the SM contributions, traditionally mediated by the $W$ boson, to include one-loop flavor-changing processes in the minimal G2HDM facilitated by new charged current interactions. These interactions are mediated by the dark Higgs ($\cal D$), the charged Higgs (${\cal H}^\pm$), and the complex dark gauge boson (${\cal W}^{p,m}$), the latter of which is a candidate for dark matter, involving both SM quarks and new heavy quarks in the loops.

We have derived new contributions to the Wilson coefficients $\Delta C_7^{(\prime)}$ and $\Delta C_8^{(\prime)}$, with our numerical results illustrated in Fig.~\ref{fig:DeltaC7C8} and Fig.~\ref{fig:DeltaC7C8p}. Within our parameters of interest, we found that the contributions from ${\cal H}^\pm$ loop diagrams significantly dominate over those from $\cal D$ and ${\cal W}^{p,m}$ loop diagrams. This dominance is due to the explicit mass factor of SM up-type quarks ( mainly top quark) in the Yukawa coupling involving the charged Higgs, SM down-type quarks, and up-type new heavy quarks, as shown in (\ref{YukChHd}), 
plus the requirement of mass insertions for internal new heavy quark lines to induce the chirality flipped magnetic and electric dipole operators in (\ref{O7O7prime}) and (\ref{O8O8prime}). Additionally, the impact of the ${\cal H}^\pm$ loop diagrams is more significant in regions with lighter masses for $m_{{\mathcal H}^\pm}$ and the new heavy up-type quarks.

Interestingly, we observed that the contributions from $\cal D$ loop and ${\cal W}^{p,m}$ loop diagrams diminish when the masses of the three generations of new heavy quarks running in the loop are degenerate. In contrast, the contributions from ${\cal H}^\pm$ loop diagrams persist under such conditions but also vanish should the masses of the three generations of SM up-type quarks are set to be degenerate as well.

Through an exhaustive parameter space scan, constrained by data from various low-energy flavor observables, we have showcased our main results in Fig.~\ref{fig:corner1}. Owing to the predominant contribution of the charged Higgs loop diagrams on flavor-changing bottom quark processes, stringent lower bounds have been placed on the masses of the particles involved in the loop, including the charged Higgs and new heavy up-type quarks. Notably, the lower bound on the charged Higgs mass is more restrictive in regions with smaller masses of the new heavy up-type quarks.

By integrating these constraints from low-energy flavor experiments with those from theoretical conditions on the scalar potential and oblique parameters, we have established lower bounds on the masses of both the charged and dark Higgs. Specifically, regions where $m_{\mathcal{H}^\pm} \lesssim 250$ GeV and $m_
\mathcal D \lesssim 100$ GeV are excluded at the $95\%$ confidence level based on our analysis.

Due to the peculiar embedding the two Higgs doublets into a two dimensional irreducible representation of the hidden $SU(2)_H$ in G2HDM, the Yukawa couplings of charged Higgs are highly correlated with the SM Higgs Yukawa couplings. Thus flavor physics is quite interesting and rich in G2HDM, as demonstrated in this work for $b \to s (\gamma,g)$ in $B$ physics, as well as in the analogous leptonic process of $l_i \to l_j \gamma$ in~\cite{Tran:2022cwh}. Many other low energy flavor physics can be explored further. For instance, 
new contributions from G2HDM to the Wilson's coefficients $
\Delta C_9^{(\prime)}$ and $\Delta C_{10}^{(\prime)}$, which are relevant to various observables for the semi-leptonic processes $B \to (X_s, K^{(*)} ) l^+ l^-, K^{(*)} \nu \bar \nu$ and may have some rooms for new physics according to the recent global fit studies in LEFT~\cite{Wen:2023pfq} and in SMEFT~\cite{Chen:2024jlj}, may be of potentially significance in providing additional constraints. Work on this direction is now in progress and will be reported elsewhere.

\section*{Acknowledgments}
This work was supported in part by the National Natural Science Foundation of China, grant Nos. 19Z103010239 and 12350410369 (VQT) and U1932104 (FRX), and the NSTC grant No. 111-2112-M-001-035 (TCY).
VQT would like to thank the Medium and High Energy Physics group at the Institute of Physics, Academia Sinica, Taiwan for their hospitality during the course of this work. TCY would like to thank Khiem Hong Phan for the hospitality he received at Duy Tân University, HCMC,  Vi\d{ê}t Nam, where the final phase of this work was completed.

\newpage

\appendix

\section{One-loop Induced Amplitudes for $b \to s \gamma$\label{app:LoopAmps}}

For the SM $W$ boson, there are only two diagrams in the unitary gauge as depicted in Fig.~\ref{fig:bsgammaW}. One obtains
\beq
\label{FFMESMW}
A^{(M,E)} \left( W \right) = A^{(M,E)}_1 \left( W \right) + A^{(M,E)}_2 \left( W \right) \; ,
\eeq
with
\begin{align}
\label{AMW1}
A^M_1 \left( W \right) & =   +  \left( \frac{g^2}{8} \right)
\sum_j \left( V_{\rm CKM} \right)^*_{j2} \left( V_{\rm CKM}\right)_{j3} \nonumber \\ 
& \;\;\;\;\;\;\;\;\;\;\;\; \times \left[ \mathcal I  \left( m_b , m_s, m_{u_j}, m_W \right)  +
\mathcal I  \left( m_b , m_s, - m_{u_j}, m_W \right)  \right] \; , \\
\label{AEW1}
A^E_1 \left( W \right) & =  -  i \left( \frac{g^2}{8} \right) 
\sum_j \left( V_{\rm CKM} \right)^*_{j2} \left( V_{\rm CKM}\right)_{j3} \nonumber \\ 
& \;\;\;\;\;\;\;\;\;\;\;\; \times \left[ \mathcal I  \left( m_b , -m_s, m_{u_j}, m_W \right)  +
\mathcal I  \left( m_b , -m_s, - m_{u_j}, m_W \right)  \right] 
\; ,
\end{align}
and
\begin{align}
\label{AMW2}
A^M_2 \left( W \right) & =   \left(-Q_u \right) \left( \frac{g^2}{8}  \right)
\sum_j \left( V_{\rm CKM} \right)^*_{j2} \left( V_{\rm CKM}\right)_{j3} \nonumber \\ 
& \;\;\;\;\;\;\;\;\;\;\;\; \times \left[ \mathcal J  \left( m_b , m_s, m_{u_j}, m_W \right)  +
\mathcal J  \left( m_b , m_s, - m_{u_j}, m_W \right)  \right] \; , \\
\label{AEW2}
A^E_2 \left( W \right) & =   \left( -  i \right) \left(-Q_u \right) \left(  \frac{g^2}{8}  \right)
\sum_j \left( V_{\rm CKM} \right)^*_{j2} \left( V_{\rm CKM}\right)_{j3} \nonumber \\ 
& \;\;\;\;\;\;\;\;\;\;\;\; \times \left[ \mathcal J  \left( m_b , -m_s, m_{u_j}, m_W \right)  +
\mathcal J  \left( m_b , -m_s, - m_{u_j}, m_W \right)  \right] 
\; .
\end{align}

For the dark Higgs $\mathcal D$ diagram in Fig.~\ref{fig:bsgammaD}, we have
\bea
\label{FFMDarkHiggs}
A^M \left( \mathcal D \right) & = & (-Q_d) 
\left[ \sum_j \left( S^{\mathcal D}_d \right)^*_{j2} \left( S^{\mathcal D}_d \right)_{j3} \mathcal K ( m_b, m_s, m_{d^H_j}, m_{\mathcal D} ) \right. \nonumber \\ 
&  & \;\;\;\;\;\;\;\;\;\;\;\;\;\;\;  + \left. \sum_j \left( P^{\mathcal D}_d \right)^*_{j2} \left( P^{\mathcal D}_d \right)_{j3} \mathcal K ( m_b, m_s, - m_{d^H_j}, m_{\mathcal D} ) \right]\; , \\
\label{FFEDarkHiggs}
A^E \left( \mathcal D \right) & = & i (-Q_d) \left[ \sum_j \left( P^{\mathcal D}_d \right)^*_{j2} \left( S^{\mathcal D}_d \right)_{j3} \mathcal K ( m_b, - m_s, m_{d^H_j}, m_{\mathcal D} ) \right. \nonumber \\ 
&  & \;\;\;\;\;\;\;\;\;\;\;\;\;\;\;  + \left. \sum_j \left( S^{\mathcal D}_d \right)^*_{j2} \left( P^{\mathcal D}_d \right)_{j3} \mathcal K ( m_b, - m_s, - m_{d^H_j}, m_{\mathcal D} ) \right]\; .
\eea

For the charged Higgs $\mathcal H^\pm$, from the two diagrams in Fig.~\ref{fig:bsgammaH}, we get
\beq
\label{FFMEChargedHiggs}
A^{(M,E)} \left( \mathcal H \right) = A^{(M,E)}_1 \left( \mathcal H \right) + A^{(M,E)}_2 \left( \mathcal H \right) \; ,
\eeq
with
\bea
\label{AMChargedHiggs1}
A^M_1 \left( \mathcal H \right) & = &  \sum_j \left( y^{\mathcal H}_u \right)^*_{j2} \left( y^{\mathcal H}_u \right)_{j3} \nonumber \\
  & & \;\;\;\;\;\;\;  \times
\left[ \mathcal L ( m_b, m_s, m_{u^H_j}, m_{\mathcal H} ) + \mathcal L ( m_b, m_s, - m_{u^H_j}, m_{\mathcal H} ) \right]  \; , \\
\label{AEChargedHiggs1}
A^E_1 \left( \mathcal H \right) & = &  -i \sum_j \left( y^{\mathcal H}_u \right)^*_{j2} \left( y^{\mathcal H}_u \right)_{j3} \nonumber \\
  & & \;\;\;\;\;\;\;  \times
\left[ \mathcal L ( m_b, - m_s, m_{u^H_j}, m_{\mathcal H} ) + \mathcal L ( m_b, - m_s, - m_{u^H_j}, m_{\mathcal H} ) \right]  \, , 
\eea
and
\bea
\label{AMChargedHiggs2}
A^M_2 \left( \mathcal H \right) & = &  \left(- Q_u  \right)
 \sum_j \left( y^{\mathcal H}_u \right)^*_{j2} \left( y^{\mathcal H}_u \right)_{j3} \nonumber \\
  & & \;\;\;\;\;\;\;  \times
 \left[ \mathcal K ( m_b, m_s, m_{u^H_j}, m_{\mathcal H} ) + \mathcal K ( m_b, m_s, - m_{u^H_j}, m_{\mathcal H} ) \right] \; , \\
\label{AEChargedHiggs2}
A^E_2 \left( \mathcal H \right) & = &  ( - i )\left(- Q_u  \right)
 \sum_j \left( y^{\mathcal H}_u \right)^*_{j2} \left( y^{\mathcal H}_u \right)_{j3} \nonumber \\
  & & \;\;\;\;\;\;\;  \times
 \left[ \mathcal K ( m_b, - m_s, m_{u^H_j}, m_{\mathcal H} ) + \mathcal K ( m_b, - m_s, - m_{u^H_j}, m_{\mathcal H} ) \right] \, .
\eea

Finally, for the contributions from the dark matter gauge boson $\mathcal W$ in the unitary gauge as depicted in Fig.~\ref{fig:bsgammaWprime}, we obtain
\bea
\label{FFMWprime}
A^M \left( \mathcal W \right) & = & (-Q_d) \left( \frac{g_H^2}{8} \right) 
 \sum_j \left( V^H_d \right)_{2j} \left( V^H_d \right)^*_{3j} \nonumber \\
 & & \;\;\;\;\;  \times
\left[ \mathcal J ( m_b, m_s, m_{d^H_j}, m_{\mathcal W} ) + \mathcal J ( m_b, m_s, - m_{d^H_j}, m_{\mathcal W} ) \right] \; , \\
\label{FFEWprime}
A^E \left( \mathcal W \right) & = & i (-Q_d) \left( \frac{g_H^2}{8} \right) 
 \sum_j \left( V^H_d \right)_{2j} \left( V^H_d \right)^*_{3j} \nonumber \\
 & & \;\;\;\;\;  \times
\left[ \mathcal J ( m_b, - m_s, m_{d^H_j}, m_{\mathcal W} ) + \mathcal J ( m_b, - m_s, - m_{d^H_j}, m_{\mathcal W} ) \right] \, .
\eea
In the above equations, $\mathcal I$, $\mathcal J$, $\mathcal K$ and $\mathcal L$ are Feynman parametrization loop integrals that can be found in the following Appendix~\ref{app:loopIntegrals}.

\newpage 

\section{Feynman Parametrization Loop Integrals\label{app:loopIntegrals}}
For convenience, we collect here the loop integrals $\mathcal{I}$, $\mathcal J$, $\mathcal K$ and $\mathcal L$ which were derived previously in~\cite{Tran:2022cwh} (See also~\cite{Lindner:2016bgg}). 
Integrals $\mathcal{I}$ and $\mathcal{J}$ entered in the vector gauge boson exchange diagrams, like those in Figs.~\ref{fig:bsgammaW} and \ref{fig:bsgammaWprime}, while $\mathcal{K}$ and $\mathcal{L}$ entered in the scalar exchange diagrams, like those in Figs.~\ref{fig:bsgammaD} and \ref{fig:bsgammaH}. 
We have kept all the external ($m_i$ and $m_j$) and internal ($m_k$ and $m_X$) masses in these integrals. 
We have checked that if the external masses are small compared with the internal ones like in the $s \to d$ transition from the $W$ exchange diagrams, series expansions of the expressions of $\mathcal{I}$ and $\mathcal{J}$ presented below can be used to reproduce the well-known SM results~\cite{Inami:1980fz} of heavy quark effects to leading order in $m_s$ and $m_d$.

To avoid word cluttering, we  denote $z \equiv 1-x-y$ in what follows.

\subsection{Integral $\mathcal I$ and $\mathcal J$}

\begin{align}
\label{Integral-I}
\mathcal I & \left( m_i , m_j , m_k , m_X \right) \nonumber \\ 
& = \int_0^1 {\rm d} x \int_0^{1-x} {\rm d} y \Biggl\{ \frac{1} { -x z m_i^2 - x y m_j^2 + x m_k^2 + (1 - x ) m_X^2 } \Biggr. \nonumber \\
& \;\;\;\; \times \Biggl[ \left( \bigl( y + 2 z \left( 1 - x \right) \bigr) + \bigl( z + 2 y \left( 1 - x \right) \bigr) \frac{ m_j }{ m_i } - \, 3  \left( 1 - x \right)  \frac{ m_k }{ m_i }  \right)
\Biggr. \nonumber \\
& \;\;\;\;\;\;\;\;\; + \Biggl. \frac{m_i^2}{m^2_{X}}  x^2 \left(   z^2  + y^2 \frac{m^3_j }{m_i^3} 
+ y z \frac{m_j}{m_i} \left( 1 + \frac{m_j}{m_i} \right)  - \frac{m_j m_k }{m_i^2} \right)  \Biggr]  \nonumber \\
&  \;\;\;\; + \frac{1}{m^2_X} \left( x (1-z)  + y + \bigl( x \left(1 - y \right) + z \bigr) \frac{m_j}{m_i}  -  \frac{m_k}{m_i} \right) \nonumber \\
&  \;\;\;\; + \frac{1}{m^2_X} \left(  2 - x \left( 3 - 4 z \right) - 3 y - z + \bigl(  2 - x \left( 3 - 4 y \right) - y - 3 z \bigr) \frac{ m_j }{ m_i }  \right) \nonumber \\
& \;\;\;\;\;\;\;\;\; \Biggl.  \times  \log \left( \frac{m^2_X} {-x z m_i^2 - x y m_j^2 + x m_k^2 + (1 - x ) m_X^2} \right)  \Biggr\} \; .
\end{align} 

We note that this integral $\mathcal I$ is for the diagram with two internal charged vector bosons $X$ coupled to the external photon 
computed using the unitary gauge. The third line of Eq.~(\ref{Integral-I}) comes from the product of the transverse pieces of the two vector boson propagators, while all the remaining terms are 
due to the product of the transverse and longitudinal pieces of these two propagators. The product of longitudinal pieces do not give rise to the contributions for the transition magnetic and 
electric dipole form factors. 


\begin{align}
\label{Integral-J}
\mathcal J & \left( m_i , m_j , m_k , m_X \right) \nonumber \\ 
& = - \int_0^1 {\rm d} x \int_0^{1-x} {\rm d} y \Biggl\{ \frac{1} { -x z m_i^2 - x y m_j^2 + (1-x) m_k^2 + x m_X^2 } \Biggr. \nonumber \\
& \;\;\;\; \times \Biggl\lceil  2 x \left( (1-z) + (1-y) \frac{m_j}{m_i} - 2  \frac{m_k}{m_i} \right) \Biggr. \nonumber \\
& \;\;\;\;\;\;\;\;\; + \frac{m_i^2}{m^2_X} \Biggl( (1-x) \left( \frac{m_j}{m_i} - \frac{m_k}{m_i} \right) 
\left( z + y \frac{m_j}{m_i} \right) \left( 1 - \frac{m_k}{m_i} \right) \Biggr.  \nonumber \\
& \;\;\;\;\;\;\;\;\;\;\;\;\;\;\;\;\;\;\;\;\;\; - z \left( \frac{m_j}{m_i} - \frac{m_k}{m_i} \right) \left(  (1-x (1 - z) ) + x y \frac{m_j^2}{m_i^2} \right) \nonumber \\
& \Biggl. \Biggl. \;\;\;\;\;\;\;\;\;\;\;\;\;\;\;\;\;\;\;\;\;\;  - \, y \left( 1 - \frac{m_k}{m_i} \right) \left( xz + (1-x (1 - y) ) \frac{m_j^2}{m_i^2} \right) \Biggr) \Biggr\rfloor \nonumber \\
&  \;\;\;\;  +  \frac{1}{m^2_X} \left( y  + z \frac{m_j}{m_i } -  \left( 1 - x  \right) \frac{m_k}{m_i} \right) \nonumber \\
&  \;\;\;\; + \frac{1}{m^2_X} \left( (1 - 3y )  + (1 - 3 z ) \frac{m_j}{m_i} +  \left( 1 - 3 \, x \right) \frac{m_k}{m_i} \right) \nonumber \\
& \;\;\;\;\;\;\;\;\; \Biggl. \times \log \left( \frac{m^2_X} {-x z m_i^2 - x y m_j^2 + (1-x) m_k^2 + x m_X^2} \right)  \Biggr\} \; .
\end{align} 
We note that this integral $\mathcal J$ is for the diagram with one internal charged or neutral gauge boson $X$ exchange while the external photon couples to the internal charged fermion. The diagram is also computed using the unitary gauge. 
The third line of Eq.~(\ref{Integral-J}) comes from the transverse piece of the vector boson propagator, while the remaining terms come entirely from the 
longitudinal piece of the propagator.  

\vfill

\subsection{Integral $\mathcal K$ and $\mathcal L$}
\begin{align}
\label{Integral-K}
\mathcal K \left( m_i , m_j , m_k , m_X \right) & = 
\int_0^1 {\rm d} x \int_0^{1-x} {\rm d} y  \nonumber \\
& \;\;\;\; \times \left[ \frac{ x \left( y +  z \frac{m_j}{m_i} \right) +  (1-x) \frac{m_k}{m_i} } { -x y m_i^2 - xz m_j^2  + (1-x) m_k^2 + x m_X^2 } \right] \; .
\end{align} 


\begin{align}
\label{Integral-L}
\mathcal L \left( m_i , m_j , m_k , m_X \right) & = -
\int_0^1 {\rm d} x \int_0^{1-x} {\rm d} y \nonumber \\
& \;\;\;\; \times \left[ \frac{ x \left( y +  z \frac{m_j}{m_i} +  \frac{m_k}{m_i} \right) } { -x y m_i^2 - xz m_j^2  + x m_k^2 + (1-x) m_X^2 } \right] \; .
\end{align} 

\end{document}